\documentclass[11pt,a4paper]{article}
\usepackage{bbm}
\pdfoutput=1
\usepackage{jheppub}
\usepackage{amsmath}
\usepackage{epsfig}
\usepackage{amssymb}
\usepackage{graphics}
\usepackage[active]{srcltx}
\usepackage{epstopdf}
\usepackage{subfigure}

\newcommand{\be}{\begin{equation}}
\newcommand{\ee}{\end{equation}}
\newcommand{\bea}{\begin{eqnarray}}
\newcommand{\eea}{\end{eqnarray}}
\newcommand{\bean}{\begin{eqnarray*}}
\newcommand{\eean}{\end{eqnarray*}}
\newcommand{\nn}{\nonumber \\}

\def\O #1{\overline{#1}}

\def\W #1{\widetilde{#1}}

\def\braket#1{\left\langle #1 \right\rangle}

\def\eref#1{(\ref{#1})}

\def\a{{\alpha}}

\def\b{{\beta}}

\def\vev{\braket}

\def\Spaa{\vev}

\def\ruleI#1{\mathcal{R}_{{\scriptsize\mbox{ule}}}^{{\tiny\mbox{I}}}[#1]}
\def\ruleII#1{\mathcal{R}_{{\scriptsize\mbox{ule}}}^{{\tiny\mbox{II}}}[#1]}
\def\ruleIII#1{\mathcal{R}_{{\scriptsize\mbox{ule}}}^{{\tiny\mbox{III}}}[#1]}

\def\Label#1{\label{#1}%
  \smash{\hbox to0pt{\raise1ex\hbox{\tiny[#1]}\hss}}}

\title{Derivation of Feynman Rules for Higher Order Poles Using Cross-ratio Identities in CHY Construction}
\author[a]{Kang Zhou,}
\author[b]{Junjie Rao,}
\author[b,c]{Bo Feng\footnote{The unusual ordering of authors instead of the
standard alphabetical one is for postdocs to get proper recognition of contributions under
the current out-dated practice in China.}}

\affiliation[a]{School of Mathematical Sciences,  Zhejiang University,\\  No.38, Zheda Road, Hangzhou, 310027, P.R. China.}
\affiliation[b]{Zhejiang Institute of Modern Physics, Department of Physics,
  Zhejiang University,\\ No.38, Zheda Road, Hangzhou, 310027, P.R. China.}
\affiliation[c]{Center of Mathematical Science,
  Zhejiang University,\\ No.38, Zheda Road, Hangzhou, 310027, P.R. China.}

\emailAdd{zkzrzm@126.com}
\emailAdd{raojunjie@zju.edu.cn}
\emailAdd{fengbo@zju.edu.cn}

\date{\today}
\abstract{In order to generalize the integration rules to general CHY integrands
which include higher order poles, algorithms are proposed in two directions. One is to conjecture new rules, and the other is to use the
cross-ratio identity method. In this paper,we use the cross-ratio identity approach
to re-derive the conjectured integration rules involving higher order poles for several special cases:
the single double pole, single triple pole and duplex-double pole. The equivalence
between the present formulas and the previously conjectured ones is discussed for the first two situations.
}


\begin{document}

\maketitle \flushbottom

\section{Introduction}
\label{secintro}

The well-known Cachazo-He-Yuan (CHY) formulation \cite{Cachazo:2013gna,Cachazo:2013hca,Cachazo:2013iea,Cachazo:2014nsa,Cachazo:2014xea}
is an elegant new representation of
tree-level amplitudes for massless particles in arbitrary dimensions, as given by
\bea
{\cal A}_n=\int{\prod_{i=1}^nd z_i\over
\mbox{vol}~SL(2,\mathbb{C})}{\prod_a}'\delta(\mathcal{E}_a)~I_n^{{\tiny\mbox{CHY}}}\,,
\eea
which possesses the M\"obius $SL(2,\mathbb{C})$ invariance. It expresses amplitudes of a large variety
of quantum field theories as multi-dimensional contour integrals over auxiliary variables
$z_i$'s, which are completely localized on the Riemann sphere by constraints known as the
scattering equations
\bea
\mathcal{E}_a\equiv\sum_{b\in\{1,2,\ldots,n\}\setminus\{a\}}~{s_{ab}\over
z_{ab}}=0\,,~~~~a=1,2,\ldots,n\,,
\eea
where $s_{ab}\equiv 2k_a\cdot k_b$ is the Mandelstam variable and $z_{ab}$
is defined as $z_{ab}\equiv z_a-z_b$. The scattering equations and the integration measure
are universal, while the integrand $I_n^{{\tiny\mbox{CHY}}}$ obeys
some general constraints (such as of weight-4 under M\"obius transformations),
and it also depends on the specific field theory.

This formulation indicates that a tree amplitude can be calculated
by solving scattering equations and summing over different
solutions. However, it is hardly possible to get direct solutions
beyond five points due to the Abel-Ruffini theorem for algebraic systems.
Thus, to search for a new computation method to avoid the explicit solutions of
algebraic equations becomes a crucial challenge. Among investigations from
various directions \cite{Kalousios:2015fya,Cardona:2015eba,Cardona:2015ouc,Dolan:2015iln,Huang:2015yka,Sogaard:2015dba,Bosma:2016ttj,
Zlotnikov:2016wtk,Cachazo:2015nwa,Gomez:2016bmv,Cardona:2016bpi,Baadsgaard:2015voa,Baadsgaard:2015ifa}, the so-called integration rule method inspired
by string theory is one of the most efficient and systematic approaches \cite{Baadsgaard:2015voa,Baadsgaard:2015ifa,Baadsgaard:2015hia}.
This approach replies only on the CHY integrands,
without mentioning the solutions of scattering equations and the integration measure.
Applying this method, one can extract all the correct pole structures from
a given integrand, and directly obtain the result via the corresponding
Feynmann diagrams, rather than solving the
scattering equations. However, one shortcomings of the original integration rules \cite{Baadsgaard:2015voa,Baadsgaard:2015ifa,Baadsgaard:2015hia} is that it requires
the CHY integrand under consideration containing simple poles only, therefore
cannot be applied to arbitrary physically acceptable integrands in general.

To handle this disadvantage, there are two alternative approaches. One is
to derive integration rules for higher order poles \cite{Huang:2016zzb}, and the other is to reduce
terms containing higher order poles into those with simple poles only \cite{Baadsgaard:2015voa,Bjerrum-Bohr:2016juj}.
In the first direction, integration rules for several special configurations
of CHY integrands with higher order poles are conjectured in \cite{Huang:2016zzb}. Although its analytic proof is absent,
these rules are numerically verified. A more hopeful
approach comes from the second direction, thanks to the discovery of the cross-ratio identities,
which reveals relations between different rational functions of $z_{ij}$'s \cite{Cardona:2016gon}.
By applying these identities iteratively, one can expand a term involving
higher order poles as terms with simple poles only.
After this decomposition, one can obtain the integrated result
through the original integration rules. The fesasibility of this algorithm is verified
in \cite{Zhou:2017mbe}. Further applications of the cross-ratio identities can be seen in \cite{Bjerrum-Bohr:2016axv,Nandan:2016pya,Huang:2017ydz}.

It is natural to ask: can we prove the conjectured integration rules of higher order poles in \cite{Huang:2016zzb}
via the cross-ratio identity method? In this paper, we will derive these rules analytically by applying
those identities. Our derivation depends on the choices of the cross-ratio identities,
thus different choices yield different expressions for the same pole configuration.
The expressions of rules in this paper will be different from
those conjectured in \cite{Huang:2016zzb}, and we will prove their equivalence for two cases.

This paper is organized as follows. In \S\ref{secprepare}, we summarize some useful notations, and some
general properties of the CHY integrands and cross-ratio identities. In \S\ref{secrule1}, we derive
the integration rule for CHY integrands containing a single double pole. In \S\ref{secrule2}, we derive
the integration rule for that containing a single triple pole. For this case, the equivalence between
our formula and the conjectured one is rather non-trivial, and the proof of their equivalence
indicates a new kind of integration rule, as will be discussed explicitly. In \S\ref{secrule3}, we
derive the integration rule for those containing duplex-double poles, regarding a simplest special case.
A brief conclusion is given in \S\ref{secconclu}.

\section{Preparation}
\label{secprepare}

Before going to the details, for clarity we give a summary of notations.
Each CHY integrand corresponds to a weight-4 graph, in which $n$ nodes are connected
by a number of lines. Each line corresponds to a factor $z_{ij}$ \footnote{
To distinguish them from lines in Feynman diagrams, we will call the latter
``Feynman lines''.}. Furthermore, since factor $z_{ij}$
can appear in both the numerator and the denominator, to distinguish them, we use a solid line
to represent $z_{ij}$ in the denominator and a dashed line to represent that in the numerator.
With this assignment, the weight-4 condition becomes that there are four lines connecting to
each node, where a solid line is counted as $+1$ and a dashed line $-1$.

For a set $\Lambda$ containing $|\Lambda|$ points of $z_i$, we call a line connecting
two points in $\Lambda$ the internal line of $\Lambda$, and a line connecting at least one point
inside $\Lambda$ and the other outside $\Lambda$ the external line of $\Lambda$. The number
of internal lines of $\Lambda$ is denoted by $\mathbb{L}[\Lambda]$, and
that of external lines by $\mathbb{E}[\Lambda]$ \footnote{Again, the number
of lines are counted as $+1$ for a solid line and $-1$ for a dashed line.}.
Furthermore, we denote the number of lines connecting two sets $\Lambda_1$ and $\Lambda_2$
as $\mathbb{L}[\Lambda_1,\Lambda_2]$.

The order of poles corresponding to the set $\Lambda$ is defined as
\bea
\chi[\Lambda]=\mathbb{L}[\Lambda]-2(|\Lambda|-1)\,.~~~~\label{order1}
\eea
For convenience we call a set
corresponding to simple poles as a ``simple set'', and similar for sets corresponding to
double and triple poles. Due to the weight-4 condition,
we have $4|\Lambda|=2\mathbb{L}[\Lambda]+\mathbb{E}[\Lambda]$, thus \eref{order1} can
be rewritten as
\bea
\chi[\Lambda]=2-{\mathbb{E}[\Lambda]\over 2}\,.~~~~\label{order2}
\eea
which will be useful later. Then, we have the following corollaries:
\begin{itemize}
\item A simple set has $4$ external lines.
\item A double set has $2$ external lines.
\item A triple set has $0$ external lines.
\item If a set $\Lambda$ contains only one point $i$, i.e., $|\Lambda|=1$, we have $\mathbb{E}[\Lambda]=4$
and thus $\chi[\Lambda]=0$ from \eref{order2}.  Although it does not contribute any simple pole,
we still call it a simple set.
\end{itemize}

A set $\Lambda$ may contain many subsets which correspond to different poles. Similar
to the definition of compatible combinations for the full set of $z_i$'s, one can
define the compatible combinations for the set $\Lambda$, and denote the sum of these combinations
as ${\cal C}[\Lambda]$. The difference is, if the full set $\Lambda$ contributes a pole, ${\cal C}[\Lambda]$
must include this pole. For example, for a set $\{1,2,3,4\}$, if the full set does
not correspond to any pole but $\{1,2,3\}$, $\{1,2\}$, $\{3,4\}$
contribute poles, then ${\cal C}[\Lambda]$ is defined as\footnote{When we write down \eref{temp-1} and
\eref{temp-2}, we have assumed that all poles are simple. For non-simple poles, further modification is needed. }
\bea
{\cal C}[\Lambda]={1\over s_{123}s_{12}}+{1\over s_{12}s_{34}}\,.~~~\label{temp-1}
\eea
However, if the full set contributes a pole, ${\cal C}[\Lambda]$ now should be
\bea
{\cal C}[\Lambda]={1\over s_{1234}}\Big({1\over s_{123}s_{12}}+{1\over s_{12}s_{34}}\Big)\,.~~~\label{temp-2}
\eea
For the set with a single point $i$, we have ${\cal C}[\{i\}]=1$.

Each CHY integrand may give a number of terms and each term can be represented by a Feynman diagram with only cubic
vertices. Except the special case of $n=3$ (i.e., there are only three external nodes), each cubic vertex
contains at least one internal propagator. At each endpoint of this internal propagator (there are two),
two branches are produced.
They can be two external nodes, or one external node and one internal propagator, or
two internal propagators. No matter in which situation, this internal propagator will
be associated with a subset $\Lambda$ and its complement $\O \Lambda$, thus we have
\bea
{\cal C}[\Lambda]=\sum_{\Spaa{\Lambda_1\Lambda_2}}{1\over s_{\Lambda}^{\chi+1}}{\cal C}[\Lambda_1]{\cal C}[\Lambda_2]\,,~~~~\label{1-1+1}
\eea
where $\Lambda_1$ and $\Lambda_2$ are two branches (two subsets with $\Lambda_1\bigcup\Lambda_2=\Lambda$)
associated with the endpoints of subset $\Lambda$.
The summation is over all correct divisions of $\Lambda_1$, $\Lambda_2$, and a special division is
denoted as $\Spaa{\Lambda_1\Lambda_2}$.

The major machinery we use in this paper is the cross ratio identity
\bea \boxed{-s_\Lambda=-s_{\O \Lambda}=\sum_{i\in \Lambda/\{p\}}\sum_{j\in \O \Lambda/\{q\}}
s_{ij}{ z_{i p}z_{jq} \over z_{i j}
z_{pq}}}~~~~\label{s-more-symmetric-I} \eea
given in \cite{Cardona:2016gon}.  Let us give some explanations of \eref{s-more-symmetric-I}:
\begin{itemize}

\item (1) $(p,q)$ is the gauge choice. Although different gauge choices give equivalent
expressions, some choices will simplify the calculation.

\item (2) We have double sums over all subsets $\Lambda$ and $\O\Lambda$.

\item (3) For each term in the sum, we have a kinematic factor $s_{ij}$. The
two denominators $z_{ij}$ and $z_{pq}$ (fixed for all terms in the sum) between subsets $\Lambda$ and
$\O \Lambda$ increase $\mathbb{E}[\Lambda]$ and $\mathbb{E}[\O \Lambda]$ by 2, thus from \eref{order2},
$\chi[\Lambda]$ and $\chi[\O \Lambda]$ are reduced by 1. Similarly, two numerators, i.e.,
$z_{ip}$ in the subset $\Lambda$ and $z_{jq}$ in $\O \Lambda$,
reduce $\mathbb{L}[\Lambda]$ and $\mathbb{L}[\O \Lambda]$ by 1, thus from \eref{order1}, $\chi[\Lambda]$ and
$\chi[\O \Lambda]$ are reduced by 1.

\end{itemize}
As when applying the cross-ratio identities, one needs to choose a gauge which includes
two points and a set corresponding to a pole, for simplicity we use  $[p,q,\Lambda]$ to denote
the gauge choice, as well as the corresponding pole.

\section{Rule I: single double pole}
\label{secrule1}

In this section, we will derive the Feynman rule I for a single double pole.
The corresponding conjectured formula in \cite{Huang:2016zzb} is given as
\bea \ruleI{p_A,p_B,p_C,p_D}={2p_A\cdot p_C+2p_B\cdot p_D\over
2s_{AB}^2}\,.~~~\label{rule1} \eea
%

\subsection{CHY configuration}

First we try to understand  CHY configurations with only one double
pole $s_\Lambda$. This means that all subsets $A$ have
$\chi[A]\leq 0$, except one subset $\Lambda$ with $\chi[\Lambda]=1$.
Furthermore, for simplicity we will assume the numerator of CHY
integrand is just $1$ \footnote{Although we will not give a
rigorous proof for CHY integrands with nontrivial numerators in this
paper, we believe Feynman rule I will be applicable to this
more general situation. In fact, when we derive the rule III, we will
meet the situation where although the numerator is not one, the same rule I has been applied
to get the correct results.}. With the assumption above, we can give some
statements of CHY configurations.

First, we have
\bea \mathbb{E}[\Lambda]=2\,,~~~~~\label{rule-I-1}\eea
which means that there are two and only two lines connecting
subset $\Lambda$ and its complement $\O \Lambda$. Now we will show that
these two lines cannot meet at the same node. Let us
assume the two lines are given by a point $a\in \Lambda$
connecting to $\O \Lambda$, then
\bea \chi[\Lambda/\{a\}]& = & \mathbb{L}[\Lambda/\{a\}]-2(
|\Lambda|-1-1)=\mathbb{L}[\Lambda]-2-2( |\Lambda|-1-1)\nn & = &
\mathbb{L}[\Lambda]-2( |\Lambda|-1)=1\,,~~~~~\label{rule-I-2}\eea
where we have used the fact that point
$a$ has four connecting lines, so there are two lines connecting
point $a$ and the subset $\Lambda/\{a\}$. Since we have
assumed that all $\chi[A]\leq 0$ with $A\neq \Lambda$,
\eref{rule-I-2} contradicts with this assumption. This
means that there are two points $a_1, a_2\in
\Lambda$ and two points $b_1, b_2\in \O \Lambda$, such that there
are one line connecting $a_1, b_1$ and the other connecting
$a_2, b_2$.  As will be discussed cautiously, when we
apply the cross-ratio identities to find the Feynman rule, the good
gauge choice is either $[a_1, b_2,\Lambda]$ or $[a_2,
b_1,\Lambda]$ \footnote{As we have remarked, the claim above, i.e.,
there are only two lines $z_{a_1b_1}$ and $z_{a_2b_2}$, has
neglected the possibility that there are nontrivial numerators in CHY
integrands, which will bring more solid lines (i.e., factors in the
denominator) connecting $\Lambda, \O \Lambda$. To deal with this
case, one can use, for example,
\bea {z_{ab} z_{dc}\over z_{ac} z_{bc}} ={ z_{ad}\over
z_{ac}}-{z_{bd}\over z_{bc}}\,,~~~\label{z-identity}\eea
to get rid of the numerator. In this paper, for simplicity we will not
discuss such more general configurations. }.

\begin{figure}[htb]
\centering \subfigure[Original integrand]{
\label{fig:rule1:A} 
\includegraphics[scale=0.7]{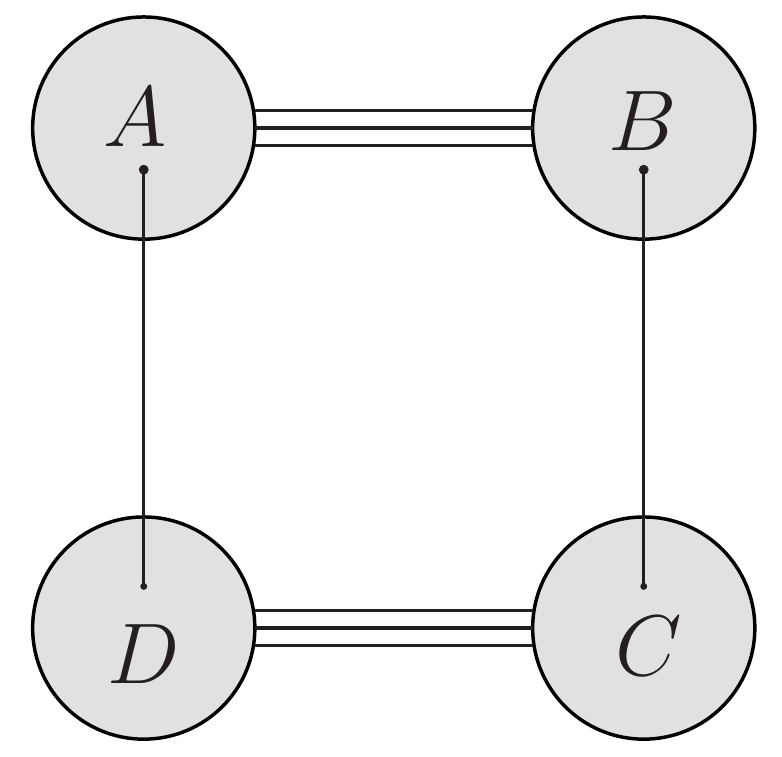}}%
\hfill%
\subfigure[Cross-ratio identity]{
\label{fig:rule1:B} 
\includegraphics[scale=0.7]{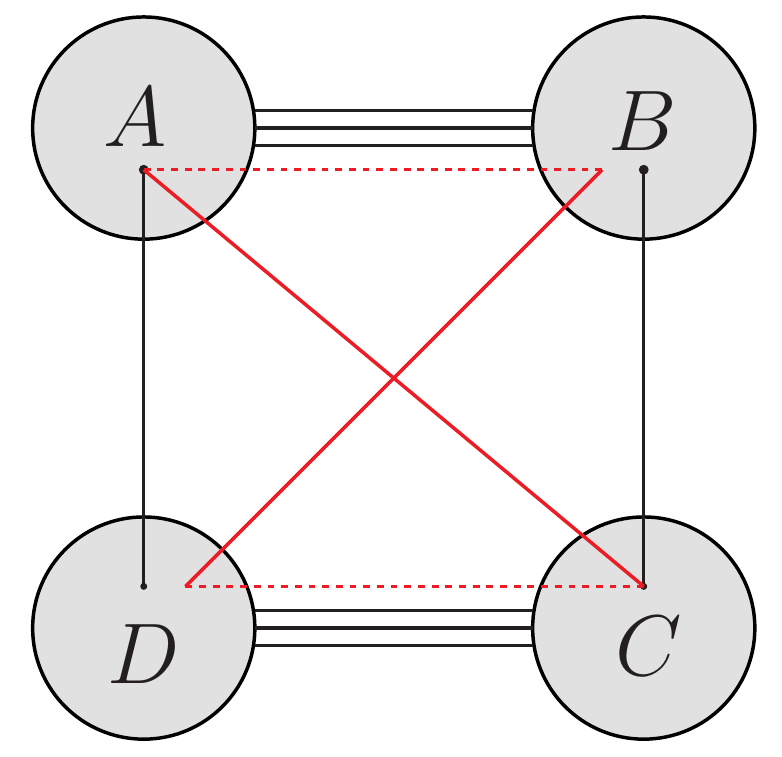}}
\caption{ Configuration of the original integrand and the cross-ratio identity of the rule I. Four red lines represent the term
${ z_{i a_0}z_{j c_0} \over z_{i j}
z_{a_0 c_0}}$
provided by the cross-ratio identity.}
\label{fig:rule1} 
\end{figure}

Second, we will show that there is no subset $A\subset \Lambda$
containing both $a_1, a_2$ satisfying $\chi[A]=0$. If such a
subset exists, we can consider its complement $\W A=\Lambda/A$. Since
$\chi[A]=0$, we have $\mathbb{E}[A]=4$. Because $a_1, a_2\in A$, we
have $\mathbb{L}[A, \W A]=2$. Now since $a_1, a_2\not\in \W A$, we
have $\mathbb{E}[\W A]=2$ so $\chi[\W A]=1$, which contradicts
with our assumption. Thus, four nodes $a_0$, $b_0$, $c_0$ and $d_0$
belong to four different simple subsets, as shown in Figure \ref{fig:rule1:A}.

Third, we will show that there is no  subset
$\Sigma=\a\bigcup\b$,  such that $\a \subset  \Lambda$, $\b\subset
\O \Lambda$ satisfying $\chi[\Sigma]=0$. If such a subset exists, we have
\bea 0 & = & \chi[\Sigma]=\mathbb{L}[\Sigma]-2(|\Sigma|-1)=
\mathbb{L}[\a]+\mathbb{L}[\b]+\mathbb{L}[\a,\b]-2(|\a|+|\b|-1)\nn
& = & (\mathbb{L}[\a]-2(|\a|-1))+(\mathbb{L}[\b]-2(|\b|-1))+
(\mathbb{L}[\a,\b]-2)\,.~~~~\label{exclude-Sigma}\eea
Since from our assumption $\chi[\a]\leq 0$ and $\chi[\b]\leq 0$, we have
 $ (\mathbb{L}[\a,\b]-2)\geq 0$. With the condition $\mathbb{L}[\a,\b]
 \leq \mathbb{L}[\Lambda,\O\Lambda]=2$,
the equation above holds when and only when
\bea \chi[\a]=0\,,~~~~\chi[\b]=0\,,~~~\mathbb{L}[\a,\b]
=2~\Longrightarrow a_1,a_2\in\a\,,~~b_1,b_2\in \b\,,\eea
which contradicts with the second observation in the previous
paragraph.

With these observations for single pole structures, we
move to the maximal compatible combinations of poles, i.e., the
possible Feynman diagrams. The claim is: {\sl all maximal
compatible combinations must contain the double subset $\Lambda$}. The reason
is: to get nonzero contributions, the number of
compatible subsets must be $n-3$. Since we have argued that there is
no single pole $\Sigma=\a\bigcup\b$ with $\a \subset \Lambda$,
$\b\subset \O \Lambda$, all single poles are either in $\Lambda$
or in $\O \Lambda$. But  $\Lambda$ can contain at most
$(|\Lambda|-2)$ compatible poles, while $\O \Lambda$ can contain at
most $(|\O\Lambda|-2)$ ones. Thus their combinations contribute
only $(n-4)$ compatible single poles, and we need to include the
double pole to get nonzero contributions.

\subsection{Some examples}

To demonstrate how to derive the Feynman rule using cross ratio
identities, we present several examples in this subsection. The same
picture will be used when derive other Feynman rules later although explicit examples
will not be given.

Let us start with the simplest example with only four points with the
CHY integrand
\bea  I_{4;a}={1\over z_{12}^3 z_{34}^3 z_{23} z_{41}}\,.\eea
This example contains only one double pole $s_\Lambda$ with
$\Lambda=\{1,2\}$. There are two lines $[14]$, $[23]$ connecting
$\Lambda$ and $\O \Lambda=\{3,4\}$. Now we apply the cross ratio
identities. There are two different gauge choices, for the first
one $[2,3, \Lambda]$, we have
\bea I_{4;a} & = & {1\over z_{12}^3 z_{34}^3 z_{23} z_{41}}\left(
{-1\over s_{12}} s_{14} {z_{43} z_{12}\over z_{14}
z_{23}}\right)={-s_{14}\over s_{12}}{1\over z_{12}^2 z_{34}^2
z_{23}^2 z_{41}^2}={-s_{14}\over s_{12}} \left( {-1\over
s_{12}}+{-1\over s_{23}}\right)={-s_{13}\over s_{12}^2}\,. \eea
For the second one $[2,4, \Lambda]$, we have
\bea I_{4;a} & = & {1\over z_{12}^3 z_{34}^3 z_{23} z_{41}}\left(
{-1\over s_{12}} s_{13} {z_{34} z_{12}\over z_{13}
z_{24}}\right)={-s_{13}\over s_{12}}{1\over z_{12}^2 z_{34}^2 z_{23}
z_{41} z_{13} z_{24}}={-s_{13}\over s_{12}} {1\over
s_{12}}={-s_{13}\over s_{12}^2}\,. \eea
We see that for the first gauge choice, pole
$s_{23}$ is introduced since in addition to original lines $z_{12}, z_{34}$,
new lines $z_{23}, z_{14}$ have been added by the
cross ratio identity. For the second gauge choice, no
new pole is introduced, so its calculation is simpler.

This phenomenon in fact suggests the general pattern. From the discussion
in the previous subsection, for a given Feynman diagram,
$\Lambda$ has been split to two subsets $A_1, A_2$ and $\O \Lambda$
has been split to two subsets $B_1, B_2$ such that $a_i\in A_i,
b_i\in B_i$, $i=1,2$. If the gauge choice is $[p\in A_2, q\in B_2,
\Lambda]$, it is possible to find two subsets $ \a \subset A_2,
\b\subset B_2$ satisfying the following conditions
\bea \chi[\a]=0\,,~~~\chi[\b]=0\,,~~~a_2\in \a\,, p\not\in \a\,;~~b_2\in \b\,,
q\not\in \b~\Longrightarrow \mathbb{L}[\a,\b]=1\,.\eea
Thus after multiplying the CHY integrand by $s_{ij} {z_{ip} z_{jq}\over z_{ij} z_{pq}}$,
we have
\bea \chi[\a\bigcup
\b]=\mathbb{L}[\a]+\mathbb{L}[\b]+\mathbb{L}[\a,\b]+1
-2(|\a|+|\b|-1)= 0\,,\eea
where the extra $+1$ comes from the denominator
$z_{ij}$ for any $i\in \a, j\in \b$. In other words, we will get a
new pole  $s_{\a\bigcup\b}$, which does not exist in the original
integrand. If we adopt the gauge  $p\in A_2, q\in B_1$, such a phenomenon
will not happen since  $\mathbb{L}[A_2, B_1]=0$. Thus to avoid this
problem, from now, we will always adopt the gauge $p\in A_2, q\in
B_1$ or $p\in A_1, q\in B_2$. Furthermore, since the splitting of
$\Lambda$ into $A_1, A_2$ is, in general, arbitrary, it is hard to
guarantee that a node $a$ always stays in $A_1$. The only exception is
that nodes $a_1, a_2$ (which connect to $b_1, b_2\in \O \Lambda$) from
the second observation in the previous subsection, i.e., $a_1$ and $a_2$
are always in different subsets of the split $\Lambda$. Thus
there is a universal gauge choice for all Feynman diagrams, which
can be either $[a_1, b_2, \Lambda]$ or $[a_2, b_1, \Lambda]$\footnote{As it will be seen in
the derivation of Feynman rule III, in general we can not find the good gauge choice such that
there is no extra pole introduced. At the same place, we will show how these extra poles have been
canceled when summing all contributions together. }.

The next example is the 5-point CHY integrand
\bea I_{5;a}=-{1\over z_{12}^3z_{23}
z_{34}^2z_{45}^2z_{53}z_{51}}\,,~~~\eea
which, according to the Feynman rule \eref{rule1}, will lead to
\bea &&{1\over s_{34}}\ruleI{\{1\},\{2\},\{3,4\},\{5\}}+{1\over
s_{45}}\ruleI{\{1\},\{2\},\{3\},\{4,5\}}\nonumber\\
&=&{1\over s_{34}}{2p_1p_{34}+2p_2p_5\over 2s_{12}^2}+{1\over
s_{45}}{2p_1p_{3}+2p_2p_{45}\over 2s_{12}^2}={s_{25}\over
s_{34}s_{12}^2}+{s_{13}\over s_{45}s_{12}^2}-{1\over
s_{12}^2}\,.~~~\Label{5p-rule-1}\eea
Now we derive this result using the cross ratio identity for the double pole $s_{12}$ (so $\Lambda=\{1,2\}$ and
$\O \Lambda=\{3,4,5\}$. The two lines connecting $\Lambda, \O \Lambda$ are $z_{15}$ and $z_{23}$. Under
the gauge $[2,5,\Lambda]$ \footnote{If we adopt the gauge $[2,3,\Lambda]$, it will produce an extra pole $s_{23}$.
If we adopt $[2,4,\Lambda]$, it will produce an extra pole $s_{15}$. Thus, to
avoid new poles, in this gauge choice $5$ is the only option if we have chosen $2$. } we have
\bea & & -{1\over z_{12}^3z_{23}
z_{34}^2z_{45}^2z_{53}z_{51}}{-1\over s_{12}} \left( s_{13} {z_{12} z_{35}\over z_{13} z_{25}} +s_{14} {z_{12} z_{45}\over z_{14} z_{25}}\right)\nn
& = & {s_{13}\over s_{12}}{z_{35}\over z_{12}^2  z_{23}
z_{34}^2z_{45}^2z_{53}z_{51}z_{13} z_{25}} +{s_{14}\over s_{12}} {z_{45}\over z_{12}^2 z_{23}
z_{34}^2z_{45}^2z_{53}z_{51}z_{14} z_{25}}\nn
& = & {s_{13}\over s_{12}} \left( {1\over s_{12} s_{34}}+{1\over s_{12} s_{45}}\right)+ {s_{14}\over s_{12}}
\left({1\over s_{12} s_{34}} \right)\nn
& = & {s_{13}+s_{14}\over s_{12}^2 s_{34}}+{s_{13}\over s_{12}^2 s_{45}}= {2 p_1\cdot p_{34}\over s_{12}^2 s_{34}}+{2 p_1\cdot p_3\over s_{12}^2 s_{45}}\,.~~~\Label{5pa-1}\eea
It seems that we get only a part of the result \eref{5p-rule-1}. Under
the gauge $[1,3,\Lambda]$ we have
\bea & & -{1\over z_{12}^3z_{23}
z_{34}^2z_{45}^2z_{53}z_{51}}{-1\over s_{12}} \left( s_{25} {z_{21} z_{53}\over z_{13} z_{25}} +s_{24} {z_{21} z_{43}\over z_{24} z_{13}}\right)\nn
& = & {s_{25}\over s_{12}}{z_{35}\over z_{12}^2  z_{23}
z_{34}^2z_{45}^2z_{53}z_{51}z_{13} z_{25}} +{s_{24}\over s_{12}} {z_{34}\over z_{12}^2 z_{23}
z_{34}^2z_{45}^2z_{53}z_{51}z_{24} z_{13}}\nn
& = & {s_{25}\over s_{12}} \left( {1\over s_{12} s_{34}}+{1\over s_{12} s_{45}}\right)+ {s_{24}\over s_{12}}
\left({1\over s_{12} s_{45}} \right)={2 p_2\cdot p_{5}\over s_{12}^2 s_{34}}+{2 p_2\cdot p_{45}\over s_{12}^2 s_{45}}\,.~~~\Label{5pa-2}\eea
Again this is not the full result but the other part of the result \eref{5p-rule-1}.
Now we see the solution: summing up \eref{5pa-1} and \eref{5pa-2}, we arrive
\bea  {  s_{13}+s_{14}+s_{25}\over 2 s_{12}^2 s_{34}}+ {s_{13}+ s_{25}+s_{24}\over 2 s_{12}^2 s_{45}}
={  2p_1\cdot p_{34}+2 p_2\cdot p_5\over 2 s_{12}^2 s_{34}}+ {2 p_1\cdot p_3+ 2p_2\cdot p_{45}\over 2 s_{12}^2 s_{45}}\,,\eea
which matches the Feynman rule \eref{5p-rule-1}.

Although simple, this example reveals that: (1) It seems that we can define different ``Feynman rules'';
(2) Different Feynman rules come from different gauge choices.

One can use more examples to better understand these two observations, as they persist to all configurations of
the current category. In the next subsection, we will give an analytic proof.

\subsection{Analytic proof}

Having understood those examples, now we can give a general analytic proof. First, from the assumption of CHY integrands,
i.e., there is one and only one subset satisfying $\chi[\Lambda]=1$ and for all others $\chi[A]\leq 0$,
we have the following statement:
\begin{itemize}

\item All maximal compatible combinations contain the subset $\Lambda$, i.e., all nonzero Feynman diagrams
contain the double pole ${1\over s^2_\Lambda}$.

\item There are four special nodes $a_0, b_0\in \Lambda$ and $c_0, d_0\in \O \Lambda$, such that
there is one line connecting nodes $a_0, d_0$ and one connecting nodes $b_0, c_0$ \footnote{Again, this claim holds only if we assume the numerator is just $1$.}.

\end{itemize}

Now we consider the Feynman rule with different gauge choices. For $[a_0, c_0, \Lambda]$, the
corresponding cross ratio identity is
\bea 1={-1\over s_\Lambda} \sum_{i\in \Lambda/\{a_0\}}\sum_{j\in \O \Lambda/\{c_0\}}
s_{ij}{ z_{i a_0}z_{j c_0} \over z_{i j}
z_{a_0 c_0}}\,,~~~~\label{ruleI-a0c0} \eea
as shown in Figure \ref{fig:rule1:B}.
Now we consider new CHY integrands ${\cal I}_{org} { z_{i a_0}z_{j c_0} \over z_{i j}
z_{a_0 c_0}}$ for each $(i,j)$ pair. First, all integrands contain only simple poles in our
construction. Second, as we have argued, under this gauge choice, all possible poles are
those already appeared in the original ${\cal I}_{org}$ and no new pole will appear. Based on these two facts,
now we focus on the contributions to a particular Feynman diagram with the pole structure ${1\over s_A s_B s_C s_D} $
(where $A\bigcup B=\Lambda$ and $C\bigcup D=\O \Lambda$,
$a_0\in A$, $b_0\in B$, $c_0\in C$ and $d_0\in D$ \footnote{As we have proven, four special nodes
must be in four different corners.}) from these CHY integrands.  For this pole structure, summation after inserting the
cross ratio identity can be divided into the following four parts:
\bea  G(a_0, c_0)_{I} & = & {-1\over s_\Lambda} \sum_{i\in A/\{a_0\}}\sum_{j\in C/\{c_0\}}
s_{ij}{ z_{i a_0}z_{j c_0} \over z_{i j}
z_{a_0 c_0}}{\cal I}_{org}\,, \nn
G(a_0, c_0)_{II} & = & {-1\over s_\Lambda} \sum_{i\in A/\{a_0\}}\sum_{j\in D}
s_{ij}{ z_{i a_0}z_{j c_0} \over z_{i j}
z_{a_0 c_0}}{\cal I}_{org}\,,\nn
G(a_0, c_0)_{III} & = & {-1\over s_\Lambda} \sum_{i\in B}\sum_{j\in C/\{c_0\}}
s_{ij}{ z_{i a_0}z_{j c_0} \over z_{i j}
z_{a_0 c_0}}{\cal I}_{org}\,,\nn
G(a_0, c_0)_{IV} & = & {-1\over s_\Lambda} \sum_{i\in B}\sum_{j\in D}
s_{ij}{ z_{i a_0}z_{j c_0} \over z_{i j}
z_{a_0 c_0}}{\cal I}_{org}\,.
\eea
Let us analyze them one by one. For $G(a_0, c_0)_{I}$, since $\chi(A)=0$ for the original integrand,
after multiplying it by ${ z_{i a_0}z_{j c_0} \over z_{i j}
z_{a_0 c_0}}$, we have
$\chi(A)=-1$ due to the numerator $z_{i a_0}$. In other words, ${\cal I}_{org} { z_{i a_0}z_{j c_0} \over z_{i j}
z_{a_0 c_0}}$ will not contain the pole ${1\over s_A}$. Similarly, the numerator $z_{j c_0}$ will lead to the fact
that there is no pole ${1\over s_C}$. Altogether, we find the $G(a_0, c_0)_{I}$ part will not
contribute to the pole structure ${1\over s_A s_B s_C s_D} $.
The same argument tells that, we can exclude the contribution from $G(a_0, c_0)_{II}$ and $G(a_0, c_0)_{III}$.
For the $G(a_0, c_0)_{IV}$ part, each term gives the same contribution\footnote{
It is easy to see that each term contributes the same pole structure, but it is hard to see that each term
gives with same sign. The sign can, in principle,  be determined by using either the method in \cite{Cachazo:2013iea} or
that in \cite{Cardona:2016gon}. The discussion of sign is too complicated
for us to give a general, simple argument. } to the particular pole structure, and we find the
total coefficient is given by
\bea {-1\over s_{AB}} \sum_{i\in B}\sum_{j\in D} s_{ij}= { 2
p_B\cdot p_D\over s_{AB}}\,.~~~\label{rule-I-Feyn-1}\eea
Summing over all possible splittings $A,B,C,D$, we get
\bea I^{{\tiny\mbox{CHY}}} & \to &  \sum_{B\subset \Lambda} \sum_{D\subset \O
\Lambda}{ 2 p_B\cdot p_D\over s^2_{\Lambda}}{\cal C}[A]{\cal
C}[B]{\cal C}[C]{\cal C}[D],\nn
& & a_0\in A=\Lambda/B\,,~b_0\in B\,,~c_0\in C=\O \Lambda/D\,,~d_0\in D\,,~~~\label{Case-I-Fey-I}\eea
which is one possible Feynman rule. From the same argument, one can see that for the gauge choice $[b_0, d_0, \Lambda]$, among four parts, only one gives nonzero contribution with the coefficient
\bea {-1\over s_{AB}} \sum_{i\in A}\sum_{j\in C} s_{ij}= { 2
p_A\cdot p_C\over s_{AB}}\,,~~~~\label{rule-I-Feyn-2}\eea
thus the other possible Feynman rule is
\bea I^{{\tiny\mbox{CHY}}} & \to &  \sum_{A\subset \Lambda} \sum_{C\subset \O
\Lambda}{ 2 p_A\cdot p_C\over s^2_{\Lambda}}{\cal C}[A]{\cal
C}[B]{\cal C}[C]{\cal C}[D]\,,\nn
& & a_0\in A\,,~b_0\in B=\Lambda/A\,,~c_0\in C\,,~d_0\in D=\O \Lambda/C\,.~~~\label{Case-I-Fey-II}\eea
Averaging over these two contributions, we get the Feynman rule
\bea I^{{\tiny\mbox{CHY}}} & \to &  \sum_{A\subset \Lambda} \sum_{C\subset \O
\Lambda}{ 2 p_A\cdot p_C+2 p_B \cdot p_D\over 2 s^2_{\Lambda}}{\cal
C}[A]{\cal C}[B]{\cal C}[C]{\cal C}[D],\nn
& & a_0\in A\,,~b_0\in B=\Lambda/A\,,~c_0\in C\,,~d_0\in D=\O \Lambda/C\,.~~~\Label{Case-I-Fey-III}\eea
which is the original Feynman rule we conjectured. We would like to emphasize that all three rules
\eref{Case-I-Fey-I}, \eref{Case-I-Fey-II} and \eref{Case-I-Fey-III} are correct, but when applying these rules,
one must stick to the same rule for all Feynman diagrams (i.e., all possible splittings of subsets $\Lambda\to A\bigcup B$ and $\O\Lambda\to C\bigcup D$) in order to get the correct final answer.

\section{Rule II: single triple pole}
\label{secrule2}

In this section, we consider   CHY configurations with only one triple
pole $s_\Lambda$. The equivalence between our formula and the conjectured one
proposed in \cite{Huang:2016zzb} is rather subtle, and it indicates a new kind of
integration rule involving quartic vertices, as will be discussed cautiously.

\subsection{CHY configuration}

The assumption of only single triple pole requires  that all subsets $A$ have
$\chi[A]\leq 0$, except one subset $\Lambda$ with $\chi[\Lambda]=2$.
Furthermore, for simplicity, we will assume the numerator of CHY
integrand is just $1$\footnote{Again, although we do not have
 the proof, we think the  Feynman rule could be applied   to the case with nontrivial
  numerators.}. With above assumption, we can make some
statements of CHY configurations.

\begin{figure}[htb]
\centering \subfigure[Original integrand]{
\label{fig:rule2:A} 
\includegraphics[scale=0.7]{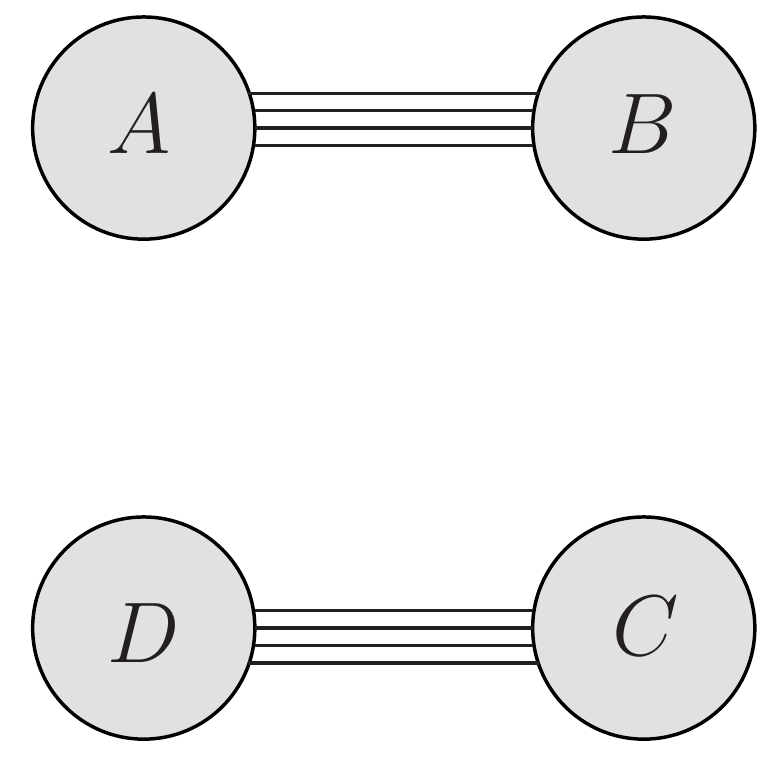}}%
\hfill%
\subfigure[Cross-ratio identity]{
\label{fig:rule2:B} 
\includegraphics[scale=0.7]{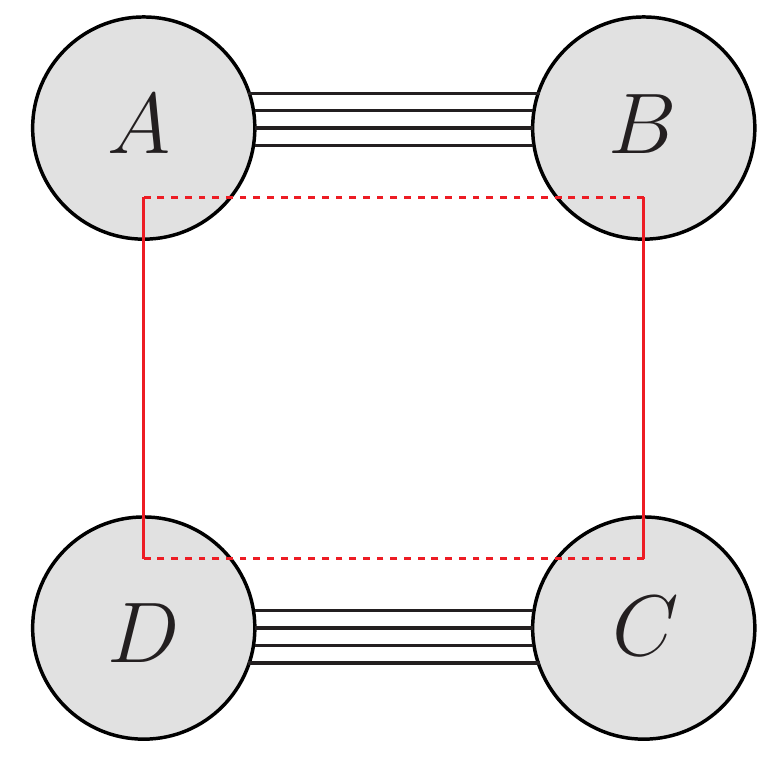}}

\caption{ Configuration of the original integrand and the cross-ratio identity of the rule  II. Four red lines represent the term
${z_{jn}z_{1i}\over
z_{ij}z_{1n}}$
provided by the cross-ratio identity. }
\label{fig:rule2} 
\end{figure}

First, the triple set $\Lambda$ satisfies $\mathbb{E}[\Lambda]=0$,
thus there is no line between $\Lambda$ and its complement subset $\O{\Lambda}$.
In other words, subset $\Lambda$ and $\O \Lambda$ give weight-4 graphes by themselves
(or a legitimate CHY integrands for smaller nodes). As a consequence,
if a subset $A\subset \Lambda$ gives a simple pole, so is $B=\Lambda\setminus A$
 since $\mathbb{E}[A]=\mathbb{E}[B]=\mathbb{L}[A,B]=4$.
Similarly, if subset $C\subset \O\Lambda$ gives a simple pole, so is  $D=\O \Lambda/C$.
To make things simpler, we will assume node $1\in \Lambda$ and node $n\in \O \Lambda$.
Furthermore, when split at the two ending points of internal propagator $s_\Lambda$,
we will assume that $1\in A$, $n\in D$ and denote the splitting as $\Spaa{ABCD}$
with $B=\Lambda/A$ and $C=\O \Lambda/D$. The corresponding configuration is shown
in Figure \ref{fig:rule2:A}.

Secondly, using \eref{exclude-Sigma}, one can observe that there is no pole corresponds to the set $\Sigma=\alpha\bigcup\beta$
such that $\alpha\subset\Lambda$ and $\beta\subset\O{\Lambda}$, since $\chi_\alpha\leq0$,
$\chi_{\beta}\leq0$ and $\mathbb{L}[\alpha,\beta]=0$. Finally,
for any maximal compatible combination,
by the same argument as in the previous section, one can see that it must contain the subset $\Lambda$.
Thus, the CHY integrand will give the contribution like
\bea \sum_{i\in A\subset \Lambda} \sum_{n\in D\subset \O \Lambda}{X\over s_{\Lambda}^3}{\cal C}[A]{\cal C}[B]{\cal C}[C]{\cal C}[D]\,,\eea
where the factor $X$ is what we need to derive for the Feynman rule.

\subsection{Derivation of new Feynman rules}

Having understood the configurations,  now we drive the rule. Since
it is the triple pole, we need to  use the cross-ratio identities
twice to reach  simple poles. At the first step, we chose the gauge
$[1,n,{\Lambda}]$ to get
\bea 1=-\sum_{i\in \Lambda\setminus\{1\}}\sum_{j\in
\O{\Lambda}\setminus\{n\}} {s_{ij}\over
s_{\Lambda}}{z_{jn}z_{i1}\over z_{ij}z_{1n}}=\sum_{i\in
\Lambda\setminus\{1\}}\sum_{j\in \bar{\Lambda}\setminus\{n\}}
{s_{ij}\over s_{\Lambda}}{z_{jn}z_{1i}\over
z_{ij}z_{1n}}\,,~~~~\label{iden1} \eea
as can be seen in Figure \ref{fig:rule2:B}.
For given $(i,j)$, the CHY integrand $I_{\rm org}{z_{jn}z_{1i}\over
z_{ij}z_{1n}}$ is nothing but the configuration we have considered
in previous section, i.e., the one with only one double pole. It is
easy to see the point: the denominators $z_{ij}$ and $z_{1n}$ have
created two connecting lines between $\Lambda$ and $\O \Lambda$. One
may worry about the numerator $z_{jn}$ in $\Lambda$ and $z_{1i}$ in
$\O\Lambda$. But as we have remarked, although our proof is given
for the case with numerator $1$, we believe the rule holds for
nontrivial numerator as current situation. Furthermore, at least for the
current CHY graph, there is always a closed Hamiltonian circle
containing $1$ and $i$ in the subset $\Lambda$ (or $n$ and $j$ in the
subset $\O \Lambda$). Thus, one can use the open-relation in
\cite{Bjerrum-Bohr:2016axv} (Eq.(3.4)) or \cite{Huang:2017ydz}
(Eq.(A12)) to eliminate the numerator (dashed line) to the case with
numerator $1$.

Since the problem has been reduced to the case with only one double pole, we can
use the previous result to write down
\bea &&I_{\rm org}{z_{jn}z_{1i}\over z_{ij}z_{1n}}  \to
\sum_{B\subset \Lambda} \sum_{D\subset \O \Lambda}{ 2 p_B\cdot
p_D\over s^2_{\Lambda}}{\cal C}[A]{\cal C}[B]{\cal C}[C]{\cal
C}[D]\,,\nn
& & 1\in A=\Lambda/B\,,~i\in B\,,~j\in C=\O \Lambda/D\,,~n\in
D\,,~~~\label{Case-II-Step2-1}\eea
for the gauge choice $[1,j,\Lambda]$ for the second step. Or
\bea &&I_{\rm org}{z_{jn}z_{1i}\over z_{ij}z_{1n}}  \to
\sum_{A\subset \Lambda} \sum_{C\subset \O \Lambda}{ 2 p_A\cdot
p_C\over s^2_{\Lambda}}{\cal C}[A]{\cal C}[B]{\cal C}[C]{\cal
C}[D]\,,\nn
& & 1\in A\,,~i\in B=\Lambda/A\,,~j\in C\,,~n\in D=\O
\Lambda/C\,,~~~\label{Case-II-Step2-2}\eea
for the gauge choice $[i,n,\Lambda]$ for the second step. Now putting
\eref{Case-II-Step2-1} and \eref{Case-II-Step2-2} back to
\eref{iden1} we get
\bea & & \sum_{i\in \Lambda\setminus\{1\}}\sum_{j\in
\O{\Lambda}\setminus\{n\}} {s_{ij}\over s_{\Lambda}}
\sum_{B\subset \Lambda} \sum_{D\subset \O \Lambda}{ 2 p_B\cdot
p_D\over s^2_{\Lambda}}{\cal C}[A]{\cal C}[B]{\cal C}[C]{\cal
C}[D]\,,\nn
& & 1\in A=\Lambda/B\,,~i\in B\,,~j\in C=\O \Lambda/D\,,~n\in
D\,,~~~\label{Case-II-Step2-1-1}\eea
for the gauge choice $[1,j,\Lambda]$. Or
\bea & & \sum_{i\in \Lambda\setminus\{1\}}\sum_{j\in
\O{\Lambda}\setminus\{n\}} {s_{ij}\over s_{\Lambda}}
\sum_{A\subset \Lambda} \sum_{C\subset \O \Lambda}{ 2 p_A\cdot
p_C\over s^2_{\Lambda}}{\cal C}[A]{\cal C}[B]{\cal C}[C]{\cal
C}[D]\,,\nn
& & 1\in A\,,~i\in B=\Lambda/A\,,~j\in C\,,~n\in D=\O
\Lambda/C\,,~~~\label{Case-II-Step2-2-1}\eea
for the gauge choice $[i,n,\Lambda]$. To continue, we
exchanging the ordering of summing to arrive at
\bea I_{\rm org} &\to&  \sum_{A\subset \Lambda} \sum_{D\subset \O
\Lambda}{(2p_B\cdot p_C) (2 p_B\cdot p_D)\over s^3_{\Lambda}}{\cal
C}[A]{\cal C}[B]{\cal C}[C]{\cal C}[D]\,,\nn
& & 1\in A\,,~~B=\Lambda/A\,,~ C=\O \Lambda/D\,,~n\in
D\,,~~~\label{Case-II-Step2-1-2}\eea
for the gauge choice $[1,j,\Lambda]$. Or
\bea I_{\rm org}& \to & \sum_{A\subset \Lambda} \sum_{D\subset \O
\Lambda}{(2p_B\cdot p_C) (2 p_A\cdot p_C)\over s^3_{\Lambda}}{\cal
C}[A]{\cal C}[B]{\cal C}[C]{\cal C}[D]\,,\nn
& & 1\in A\,,~~B=\Lambda/A\,,~ C=\O \Lambda/D\,,~n\in
D\,,~~~\label{Case-II-Step2-2-2}\eea
for the gauge choice$[i,n,\Lambda]$. Results
\eref{Case-II-Step2-1-2} and \eref{Case-II-Step2-2-2} are, in fact,
two possible Feynman rules for the triple pole\footnote{Again, we need to stick to the same
rule for all splittings of $\Spaa{ABCD}$ to get the right result.}.
It is worth to emphasize that when exchanging the ordering of the
sum, the $\sum_{i\in \Lambda\setminus\{1\}}\sum_{j\in
\O{\Lambda}\setminus\{n\}} s_{ij}$ produces $(2p_B\cdot p_C)$.
Also, changing of summation ordering is allowed because we have fixed the
first gauge choice $1\in \Lambda, n\in \O \Lambda$ for all
splittings of $\Spaa{ABCD}$. The first gauge choice is crucial for
the previous Feynman rule, while the second gauge has some natural
choice. Since we will use different gauge choices, we will use
$[1,n;1;\Lambda]$ for the gauge choice leading to the rule
\eref{Case-II-Step2-1-2} and $[1,n;n;\Lambda]$ for the gauge choice
leading to the rule \eref{Case-II-Step2-2-2}.

\subsection{Comparison with conjectured formula}

The Feynman rules found in the previous subsection  is not the one
conjectured in \cite{Huang:2016zzb}, which is given by
\bea&&\ruleII{p_A,p_B,p_C,p_D}~~~~\Label{rule2}\\
&=&{(2p_A\cdot p_C)(2p_A\cdot p_D)+(2p_B\cdot p_C)(2p_B\cdot p_D)+(2p_C\cdot p_A)(2p_C\cdot p_B)+(2p_D\cdot p_A)(2p_D\cdot p_B)\over
4s_{AB}^3}\nonumber\\
&&-{(p_A^2-p_B^2)^2+(p_C^2-p_D^2)^2\over 4s_{AB}^3}+{2\over
9}{(p_A^2+p_B^2)(p_C^2+p_D^2)\over 4s_{AB}^3}\,.~~~\nonumber\eea
Comparing
these different Feynman rules, we see that the major difference is
that the rule \eref{rule2} is gauge independent, while
rules \eref{Case-II-Step2-1-2} and \eref{Case-II-Step2-2-2} depend on both
the first gauge choice and the second gauge choice. Because of this,
the later two Feynman rules are simpler than the first one. In this
subsection, we discuss how to arrive at \eref{rule2} from
\eref{Case-II-Step2-1-2} and \eref{Case-II-Step2-2-2}.

For simplicity, let us assume that $\Lambda=\{1,2,\ldots,m\}$,
$\O{\Lambda}=\{m+1,m+2,\ldots,n\}$. The final result of a given
CHY integrand in our case will be the sum of different splittings
$\Spaa{ABCD}$
\bea {\cal A}=\sum_{1\in A\subset \Lambda}~~\sum_{n\in D\subset\O \Lambda}
~~{X\over s^3_\Lambda}{\cal C}[A]{\cal C}[B]{\cal C}[C]{\cal
C}[D]\,,~~~\label{Comp-Huang-1}\eea
with $B=\Lambda/A$ and $C=\O\Lambda/D$, where to fix the ambiguity,
we have set the subset containing node $1$ as $A$, and the subset
containing node $n$ as $D$. The factor $4$ in the denominator
\eref{rule2} implies that we should average over four different
gauge choices, just like \eref{rule1} is reproduced by average two
different gauge choices in previous sections.

Now we consider the following four different gauge choices. For the
first gauge choice $[1,n;1;\Lambda]$, we get
\bea \sum_{1\in A\subset \Lambda}~~\sum_{n\in D\subset\O \Lambda}
~~{(2p_B\cdot p_C) (2 p_B\cdot p_D)\over s^3_{\Lambda}}{\cal
C}[A]{\cal C}[B]{\cal C}[C]{\cal C}[D]\,,~~~\label{Comp-Huang-2-1}\eea
where the rule \eref{Case-II-Step2-1-2} has been used. For the
second  gauge choice $[1,n;n;\Lambda]$, we get
\bea \sum_{1\in A\subset \Lambda}~~\sum_{n\in D\subset\O \Lambda}
~~{(2p_B\cdot p_C) (2 p_A\cdot p_C)\over s^3_{\Lambda}}{\cal
C}[A]{\cal C}[B]{\cal C}[C]{\cal C}[D]\,,~~~\label{Comp-Huang-2-2}\eea
where the rule \eref{Case-II-Step2-2-2} has been used. For the third
gauge choice $[m;n;m;\Lambda]$, the thing is a little bit
complicated: since we have fixed the subset $A$ to be the one
containing node $1$, we need to consider two different cases. When
$m$ is also in the subset $A$, we get
\bea \sum_{1,m\in A\subset \Lambda}~~\sum_{n\in D\subset\O \Lambda}
~~{(2p_B\cdot p_C) (2 p_B\cdot p_D)\over s^3_{\Lambda}}{\cal
C}[A]{\cal C}[B]{\cal C}[C]{\cal
C}[D]\,.~~~\label{Comp-Huang-2-3-1}\eea
When $m$ is not in the subset $A$, we get
\bea \sum_{1\in A\subset \Lambda;m\not \in A}~~\sum_{n\in D\subset\O
\Lambda} ~~{(2p_A\cdot p_C) (2 p_A\cdot p_D)\over
s^3_{\Lambda}}{\cal C}[A]{\cal C}[B]{\cal C}[C]{\cal
C}[D]\,.~~~\label{Comp-Huang-2-3-2}\eea
Adding these two parts \eref{Comp-Huang-2-3-1} and
\eref{Comp-Huang-2-3-2} together, we get the expression for the
third gauge choice
\bea & & \sum_{1\in A\subset \Lambda}~~\sum_{n\in D\subset\O
\Lambda} ~~{(2p_A\cdot p_C) (2 p_A\cdot p_D)\over
s^3_{\Lambda}}{\cal C}[A]{\cal C}[B]{\cal C}[C]{\cal
C}[D]\nn
&+ & \sum_{1,m\in A\subset \Lambda}~~\sum_{n\in D\subset\O \Lambda}
~~{(2p_B\cdot p_C) (2 p_B\cdot p_D)-(2p_A\cdot p_C) (2 p_A\cdot
p_D)\over s^3_{\Lambda}}{\cal C}[A]{\cal C}[B]{\cal C}[C]{\cal
C}[D]\,.~~~\label{Comp-Huang-2-3}\eea
For the fourth gauge choice $[1,m+1;m+1;\Lambda]$, we need to
consider two different cases too. When $(m+1)\in D$, we get
\bea \sum_{1\in A\subset \Lambda}~~\sum_{n,(m+1)\in D\subset\O
\Lambda} ~~{(2p_B\cdot p_C) (2 p_A\cdot p_C)\over
s^3_{\Lambda}}{\cal C}[A]{\cal C}[B]{\cal C}[C]{\cal
C}[D]\,.~~~\label{Comp-Huang-2-4-1}\eea
When $(m+1)\not\in D$, we get
\bea \sum_{1\in A\subset \Lambda}~~\sum_{n\in D\subset\O
\Lambda;(m+1)\not\in D} ~~{(2p_D\cdot p_A) (2 p_D\cdot p_B)\over
s^3_{\Lambda}}{\cal C}[A]{\cal C}[B]{\cal C}[C]{\cal
C}[D]\,.~~~\label{Comp-Huang-2-4-2}\eea
Summing over \eref{Comp-Huang-2-4-1} and \eref{Comp-Huang-2-4-2},
the expression of the fourth gauge choice is
\bea & & \sum_{1\in A\subset \Lambda}~~\sum_{n\in D\subset\O
\Lambda} ~~{(2p_D\cdot p_A) (2 p_D\cdot p_B)\over
s^3_{\Lambda}}{\cal C}[A]{\cal C}[B]{\cal C}[C]{\cal C}[D]\nn
& + &\sum_{1\in A\subset \Lambda}~~\sum_{n,(m+1)\in D\subset\O
\Lambda} ~~{(2p_B\cdot p_C) (2 p_A\cdot p_C)-(2p_D\cdot p_A) (2
p_D\cdot p_B)\over s^3_{\Lambda}}{\cal C}[A]{\cal C}[B]{\cal
C}[C]{\cal C}[D]\,. ~~~\label{Comp-Huang-2-4}\eea
Now we average four different gauge choices \eref{Comp-Huang-2-1},
\eref{Comp-Huang-2-2}, \eref{Comp-Huang-2-3} and
\eref{Comp-Huang-2-4} to reach $X$ in \eref{Comp-Huang-1} as
\bea {\cal A} & = &  \sum_{\Spaa{ABCD}} \O{X}{\cal C}[A]{\cal C}[B]{\cal
C}[C]{\cal C}[D]\nn
& + &\sum_{1\in A\subset \Lambda}~~\sum_{n,(m+1)\in D\subset\O
\Lambda} ~~{(2p_B\cdot p_C) (2 p_A\cdot p_C)-(2p_D\cdot p_A) (2
p_D\cdot p_B)\over 4}{\cal C}[A]{\cal C}[B]{\cal
C}[C]{\cal C}[D] \nn
&+ & \sum_{1,m\in A\subset \Lambda}~~\sum_{n\in D\subset\O \Lambda}
~~{(2p_B\cdot p_C) (2 p_B\cdot p_D)-(2p_A\cdot p_C) (2 p_A\cdot
p_D)\over 4}{\cal C}[A]{\cal C}[B]{\cal
C}[C]{\cal C}[D]\,,~~~\label{Temp-X} \eea
where
\bea \O X= {(2p_A\cdot p_C) (2 p_A\cdot p_D)+(2p_B\cdot p_C) (2
p_B\cdot p_D)+(2p_C\cdot p_A) (2 p_C\cdot p_B)+(2p_D\cdot p_A) (2
p_D\cdot p_B)\over 4 }\,. ~~~~\label{xbar}\eea
Thus to reproduce the rule \eref{rule2}, we need to show that
\bea & & \sum_{1\in A\subset \Lambda}~~\sum_{n\in D\subset\O \Lambda}
~~-{(p_A^2-p_B^2)^2+(p_C^2-p_D^2)^2+{2\over
9}(p_A^2+p_B^2)(p_C^2+p_D^2)\over 4 s^3_\Lambda}{\cal C}[A]{\cal C}[B]{\cal C}[C]{\cal
C}[D]\nn
& = &\sum_{1\in A\subset \Lambda}~~\sum_{n,(m+1)\in D\subset\O
\Lambda} ~~{(2p_B\cdot p_C) (2 p_A\cdot p_C)-(2p_D\cdot p_A) (2
p_D\cdot p_B)\over 4s^3_\Lambda}{\cal C}[A]{\cal C}[B]{\cal C}[C]{\cal
C}[D] \nn
&+ & \sum_{1,m\in A\subset \Lambda}~~\sum_{n\in D\subset\O \Lambda}
~~{(2p_B\cdot p_C) (2 p_B\cdot p_D)-(2p_A\cdot p_C) (2 p_A\cdot
p_D)\over 4s^3_\Lambda}{\cal C}[A]{\cal C}[B]{\cal C}[C]{\cal
C}[D]\,.~~~\label{Middle-Identity} \eea
%
To make the comparison explicitly, we rewrite $(2p_B\cdot p_C) (2 p_B\cdot p_D)-(2p_A\cdot p_C) (2 p_A\cdot p_D)$
as follows,
\bea
& &(2p_B\cdot p_C) (2 p_B\cdot p_D)-(2p_A\cdot p_C) (2 p_A\cdot p_D)\nn
&=&(2p_B\cdot p_C) (2 p_B\cdot p_D)-(2p_{\Lambda}\cdot p_C) (2 p_{\Lambda}\cdot p_D)+(2p_B\cdot p_C) (2 p_{\Lambda}\cdot p_D)
+(2p_{\Lambda}\cdot p_C) (2 p_B\cdot p_D)-(2p_B\cdot p_C) (2 p_B\cdot p_D)\nn
&=&-(2p_{\overline{\Lambda}}\cdot p_C) (2 p_{\overline{\Lambda}}\cdot p_D)-(2p_B\cdot p_C) (2 p_{\overline{\Lambda}}\cdot p_D)
-(2p_{\overline{\Lambda}}\cdot p_C) (2 p_B\cdot p_D)\nn
&=&-(2s_C+2p_C\cdot p_D) (2s_D+2p_C\cdot p_D)-(2p_B\cdot p_C) (2s_D+2p_C\cdot p_D)
-(2 p_B\cdot p_D)(2s_C+2p_C\cdot p_D) \nn
&=&-(2p_C\cdot p_D)(s_C+s_D+2p_C\cdot p_D+2p_B\cdot p_C+2p_B\cdot p_D)-4s_Cs_D-s_C(2p_C\cdot p_D+4p_B\cdot p_D)\nn
& &-s_D(2p_C\cdot p_D+4p_B\cdot p_C)\nn
&=&-(2p_C\cdot p_D)(s_A-s_B)-4s_Cs_D-s_C(2p_C\cdot p_D+4p_B\cdot p_D)-s_D(2p_C\cdot p_D+4p_B\cdot p_C)\nn
&=&-s_{\Lambda}(s_A-s_B)+{\cal K}_1\,,~~~\label{Rewrite-1}
\eea
where
\bea
{\cal K}_1=-4s_Cs_D-s_C(2p_C\cdot p_D+4p_B\cdot p_D+s_B-s_A)-s_D(2p_C\cdot p_D+4p_B\cdot p_C+s_B-s_A)\,,
\eea
and similarly
\bea
& &(2p_A\cdot p_C)(2p_B\cdot p_C)-(2p_A\cdot p_D)(2p_B\cdot p_D)\nn
&=&-s_{\O\Lambda}(s_D-s_C)+{\cal K}_2\,,~~~\label{Rewrite-2}
\eea
where
\bea
{\cal K}_2=-4s_As_B-s_A(2p_A\cdot p_B+4p_C\cdot p_B+s_C-s_D)-s_B(2p_A\cdot p_B+4p_C\cdot p_A+s_C-s_D)\,.
\eea
We will prove that
\bea
\sum_{1,m\in A\subset \Lambda}~s_{\Lambda}(s_A-s_B){\cal C}[A]{\cal C}[B]
 = \sum_{1\in A\subset \Lambda}~(s_A-s_B)^2{\cal C}[A]{\cal C}[B]\,,~~~~\label{rela-1}
\eea
\bea
\sum_{m+1,n\in D\subset \Lambda}~~s_{\O\Lambda}(s_D-s_C){\cal C}[C]{\cal C}[D]
 = \sum_{n\in D\subset \Lambda}~~(s_C-s_D)^2{\cal C}[C]{\cal C}[D]\,,~~~~\label{rela-2}
\eea
and
\bea
& & \sum_{1,m\in A\subset \Lambda}~~\sum_{n\in D\subset\O \Lambda}~~{\cal K}_1{\cal C}[A]{\cal C}[B]{\cal C}[C]{\cal C}[D] \nn
&=&\sum_{1\in A\subset \Lambda}~~\sum_{n,(m+1)\in D\subset\O
\Lambda}~~{\cal K}_2{\cal C}[A]{\cal C}[B]{\cal C}[C]{\cal C}[D]\nn
&=&\sum_{1\in A\subset \Lambda}~~\sum_{n\in D\subset\O \Lambda}~~ {1\over 9}(s_A+s_B)(s_C+s_D){\cal C}[A]{\cal C}[B]{\cal C}[C]{\cal C}[D]\,.~~~~\label{rela-3}
\eea
Putting \eref{rela-1}, \eref{rela-2} and \eref{rela-3} back to \eref{Middle-Identity}, we see the identity is proved.

\begin{figure}[htb]
\centering
  \includegraphics[scale=0.7]{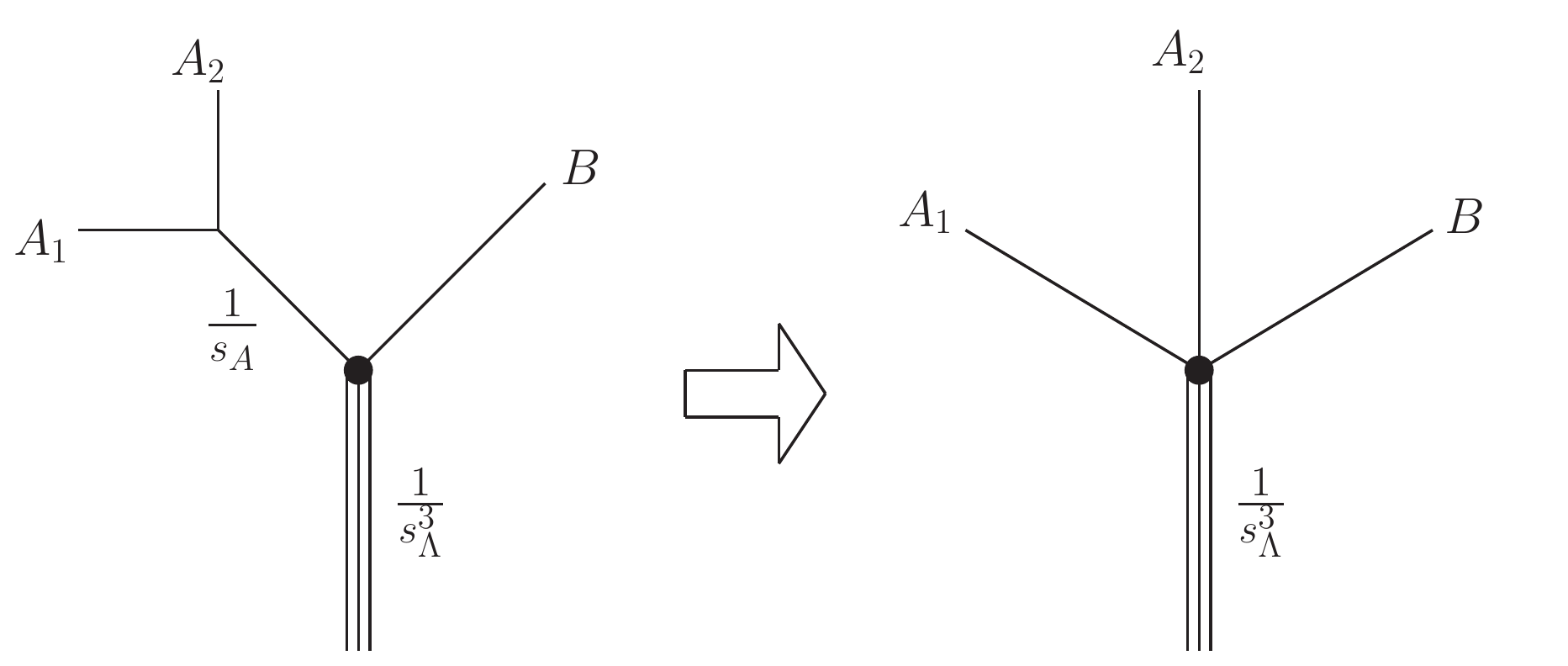}\\
  \caption{Creation of the quartic vertex from a cubic vertex.}\label{3to4}
\end{figure}

Let us start with the relation \eref{rela-1}. At the left handed side, since we have assumed that $A$ contains at least two points $1$ and $m$,
using \eref{1-1+1} we know that $s_A{\cal C}[A]{\cal C}[B]$ will remove the single pole ${1\over s_A}$ inside ${\cal C}[A]$
and split to $\sum_{\Spaa{A_1A_2}}{\cal C}[A_1]{\cal C}[A_2]$ with $A=A_1\bigcup A_2$.
This procedure removes the propagator $1/s_A$ from the cubic vertex, and creates a new quartic vertex by pushing
$A_1$ and $A_2$ to the original cubic vertex, as shown in Figure \ref{3to4}.
Similarly, if  $B$ also contains at least two points,  $s_B{\cal C}[B]=\sum_{\Spaa{B_1B_2}}{\cal C}[B_1]{\cal C}[B_2]$ with $B=B_1\bigcup B_2$. In other words, we will have
\bea
(s_A-s_B){\cal C}[A]{\cal C}[B]&=&\sum_{\Spaa{A_1A_2}}{\cal C}[A_1]{\cal C}[A_2]{\cal C}[B]-\sum_{\Spaa{B_1B_2}}{\cal C}[A]{\cal C}[B_1]{\cal C}[B_2]\,.
\eea
where to fix the ambiguity we will assume that $1\in A_1$. The summation in the first term
is over all correct divisions of $A_1$ and $A_2$, and the summation in the second term
is over all correct divisions of $B_1$ and $B_2$. We emphasize that a correct division must ensure that $A_1$, $A_2$, $B_1$ and
$B_2$ are simple sets since they corresponding to Feynman lines in Feynman diagrams.

If $B$ contains only one point, we have
\bea
(s_A-s_B){\cal C}[A]{\cal C}[B]&=&\sum_{\Spaa{A_1A_2}}{\cal C}[A_1]{\cal C}[A_2]{\cal C}[B]\,.~~~\label{sB0}
\eea
The discussion above implies that $\sum(s_A-s_B){\cal C}[A]{\cal
C}[B]$ with $1,m\in A$ will be the summation of several ${\cal
C}[\Lambda_1]{\cal C}[\Lambda_2]{\cal C}[\Lambda_3]$'s, where
$\Lambda_1$, $\Lambda_2$ and $\Lambda_3$ are simple sets satisfy
$\Lambda_1\bigcup \Lambda_2\bigcup \Lambda_3=\Lambda$. We fix the
configuration as $1\in \Lambda_1$. Thus we can consider the
coefficient of ${\cal C}[\Lambda_1]{\cal C}[\Lambda_2]{\cal
C}[\Lambda_3]$ under  a special division
$\Spaa{\Lambda_1\Lambda_2\Lambda_3}$. There are several cases:
\begin{itemize}

\item (1) If $m\in \Lambda_1$, there are three correct configurations
which are $\{A_1=\Lambda_1,A_2=\Lambda_2,B=\Lambda_3\}$,
$\{A_1=\Lambda_1,A_2=\Lambda_3,B=\Lambda_2\}$ and
$\{A=\Lambda_1,B_1=\Lambda_2,B_2=\Lambda_3\}$, thus, after
summing over all correct divisions, the coefficient of ${\cal
C}[\Lambda_1]{\cal C}[\Lambda_2]{\cal C}[\Lambda_3]$ is
$1+1-1=1$. Notice that for the third configuration above,
$\{A=\Lambda_1,B_1=\Lambda_3,B_2=\Lambda_2\}$ gives the same
situation $B=\Lambda_2\bigcup\Lambda_3$ as
$\{A=\Lambda_1,B_1=\Lambda_2,B_2=\Lambda_3\}$, so we chose only
one of them. An alternative way is to add them together and divide
by the symmetry factor $2$.

\item (2) If $m\in \Lambda_2$ ($m\in \Lambda_3$ gives the same
situation), there is only one correct configuration
$\{A_1=\Lambda_1,A_2=\Lambda_2,B=\Lambda_3\}$, thus the
coefficient is $1$.

\end{itemize}
Thus, the coefficient of any ${\cal
C}[\Lambda_1]{\cal C}[\Lambda_2]{\cal C}[\Lambda_3]$ is 1, we
get
\bea \sum_{1,m\in A\subset \Lambda}~s_{\Lambda}(s_A-s_B){\cal
C}[A]{\cal C}[B]
=s_{\Lambda}\sum_{\Spaa{\Lambda_1\Lambda_2\Lambda_3}}~{\cal
C}[\Lambda_1]{\cal C}[\Lambda_2]{\cal
C}[\Lambda_3]\,.~~~\label{ccc} \eea

One may observe that \eref{sB0} can be ignored when counting the coefficient
of ${\cal C}[\Lambda_1]{\cal C}[\Lambda_2]{\cal C}[\Lambda_3]$. The reason is, if we consider the configuration
$\{A=\Lambda_1,B_1=\Lambda_2,B_2=\Lambda_3\}$, we have assumed that $B$ contains at least two points. If we consider
other configurations, the structure of $B$ is not important. The set $\Lambda$ can always be
divided into three simple subsets if it contains at least three points. We can take the
perspective that $A$ and $B$ are constructed by three subsets, then the irrelevant cases
(such as $B$ can not be divided into two simple subsets) will be neglected automatically.

Having considered the left handed side of relation \eref{rela-1}, we
move to the right handed side, i.e., the $(s_A-s_B)^2{\cal
C}[A]{\cal C}[B]$ part. By similar argument, we can write it as
\bea (s_A-s_B)^2{\cal C}[A]{\cal
C}[B]=(s_A-s_B)\Big(\sum_{\Spaa{A_1A_2}}{\cal C}[A_1]{\cal
C}[A_2]{\cal C}[B]-\sum_{\Spaa{B_1B_2}}{\cal C}[A]{\cal C}[B_1]{\cal
C}[B_2]\Big)\,, \eea
If one of $A$ and $B$ is a single point, one term in the bracket
vanishes. Thus $\sum(s_A-s_B)^2{\cal C}[A]{\cal C}[B]$ also provides
several ${\cal C}[\Lambda_1]{\cal C}[\Lambda_2]{\cal
C}[\Lambda_3]$'s. Let us count the coefficient of ${\cal
C}[\Lambda_1]{\cal C}[\Lambda_2]{\cal C}[\Lambda_3]$ with $1\in
\Lambda_1$, under the summation $\sum(s_A-s_B)^2{\cal C}[A]{\cal
C}[B]$ which is over all divisions $\Spaa{AB}$, without the
constraint $m\in A$. There are three correct configurations,
$\{A_1=\Lambda_1,A_2=\Lambda_2,B=\Lambda_3\}$,
$\{A_1=\Lambda_1,A_2=\Lambda_3,B=\Lambda_2\}$ and
$\{A=\Lambda_1,B_1=\Lambda_2,B_2=\Lambda_3\}$, then the coefficient
is
\bea (s_{\Lambda_1 \Lambda_2}-s_{\Lambda_3})+(s_{\Lambda_1
\Lambda_3}-s_{\Lambda_2})-(s_{\Lambda_1}-s_{\Lambda_2
\Lambda_3})=s_{\Lambda_1\Lambda_2 \Lambda_3}=s_{\Lambda}\,.
\eea
Notice that if some of $\Lambda_i$'s are single points, i.e.,
$s_{\Lambda_i}=0$, the above relation still holds, as can be
verified directly. Thus, after summing over contributions from
different divisions $\Spaa{AB}$, We get
\bea \sum_{1,m\in A\subset \Lambda}~s_{\Lambda}(s_A-s_B){\cal
C}[A]{\cal C}[B]
 = \sum_{1\in A\subset \Lambda}~(s_A-s_B)^2{\cal C}[A]{\cal C}[B]=s_{\Lambda}\sum_{\Spaa{\Lambda_1\Lambda_2\Lambda_3}}~{\cal C}[\Lambda_1]{\cal C}[\Lambda_2]{\cal C}[\Lambda_3]\,,
\eea
which has proven the relation \eref{rela-1}.

Exactly similar argument shows
\bea \sum_{m+1,n\in D\subset \Lambda}~~s_{\O\Lambda}(s_D-s_C){\cal
C}[C]{\cal C}[D]
 = \sum_{n\in D\subset \Lambda}~~(s_C-s_D)^2{\cal C}[C]{\cal C}[D]=s_{\O{\Lambda}}\sum_{\Spaa{\O\Lambda_1\O\Lambda_2\O\Lambda_3}}~{\cal C}[\O\Lambda_1]{\cal C}[\O\Lambda_2]{\cal C}[\O\Lambda_3]\,,
\eea
where $\O\Lambda_1$, $\O\Lambda_2$ and $\O\Lambda_3$ are three
simple sets with $\O\Lambda_1\bigcup \O\Lambda_2\bigcup
\O\Lambda_3=\O\Lambda$, which has proved the relation \eref{rela-2}.

Then we turn to the most difficult relation \eref{rela-3}. We divide
$\sum {\cal K}_1{\cal C}[A]{\cal C}[B]{\cal C}[C]{\cal C}[D]$ into
following three parts:
\bea Y_1=\sum_{1,m\in A\subset \Lambda}~\sum_{n\in D\subset\O
\Lambda}~-(s_C+s_D)(s_B-s_A){\cal C}[A]{\cal C}[B]{\cal C}[C]{\cal
C}[D]\,, \eea
\bea Y_2=\sum_{1,m\in A\subset \Lambda}~\sum_{n\in D\subset\O
\Lambda}~-\Big(s_C(2p_C\cdot p_D+4p_B\cdot p_D)+s_D(2p_C\cdot
p_D+4p_B\cdot p_C)\Big) {\cal C}[A]{\cal C}[B]{\cal C}[C]{\cal
C}[D]\,, \eea
\bea Y_3=\sum_{1,m\in A\subset \Lambda}~\sum_{n\in D\subset\O
\Lambda}~-4s_Cs_D{\cal C}[A]{\cal C}[B]{\cal C}[C]{\cal C}[D]\,.
\eea
and will  treat them one by one.

First, using \eref{ccc} we get
\bea Y_1&=&\sum_{n\in D\subset\O \Lambda}~(s_C+s_D){\cal C}[C]{\cal
C}[D]\sum_{\Spaa{\Lambda_1\Lambda_2\Lambda_3}}~{\cal
C}[\Lambda_1]{\cal C}[\Lambda_2]{\cal C}[\Lambda_3]\nn &=&\sum_{n\in
D\subset\O \Lambda}~\Big(\sum_{\Spaa{C_1C_2}}{\cal C}[C_1]{\cal
C}[C_2]{\cal C}[D]+\sum_{\Spaa{D_1D_2}}{\cal C}[C]{\cal C}[D_1]{\cal
C}[D_2]\Big)\sum_{\Spaa{\Lambda_1\Lambda_2\Lambda_3}}~{\cal
C}[\Lambda_1]{\cal C}[\Lambda_2]{\cal C}[\Lambda_3]\,. \eea
We assume that $n\in D_1$. We can consider the coefficient of ${\cal
C}[\Lambda_1]{\cal C}[\Lambda_2]{\cal C}[\Lambda_3]{\cal
C}[\O\Lambda_1]{\cal C}[\O\Lambda_2]{\cal C}[\O\Lambda_2]$ with
$n\in\O\Lambda_1$. There are three correct configurations which are
$\{D_1=\O\Lambda_1,D_2=\O\Lambda_2,C=\O\Lambda_3\}$,
$\{D_1=\O\Lambda_1,D_2=\O\Lambda_3,D=\O\Lambda_2\}$ and
$\{D=\O\Lambda_1,C_1=\O\Lambda_2,C_2=\O\Lambda_3\}$, thus we have
\bea
Y_1=3\sum_{\Spaa{\Lambda_1\Lambda_2\Lambda_3}}\sum_{\Spaa{\O\Lambda_1\O\Lambda_2\O\Lambda_3}}
~~\big({\cal C}[\Lambda_1]{\cal C}[\Lambda_2]{\cal
C}[\Lambda_3]\big) \big({\cal C}[\O\Lambda_1]{\cal
C}[\O\Lambda_2]{\cal C}[\O\Lambda_3]\big)\,. \eea

Secondly, considering $Y_2$ gives
\bea Y_2&=&\sum_{1,m\in A\subset \Lambda}~\sum_{n\in D\subset\O
\Lambda}~-\Big((2p_C\cdot p_D+4p_B\cdot
p_D)\sum_{\Spaa{C_1C_2}}{\cal C}[C_1]{\cal C}[C_2]{\cal C}[D]\nn &
&+(2p_C\cdot p_D+4p_B\cdot p_C){\sum_{\Spaa{D_1D_2}}\cal C}[C]{\cal
C}[D_1]{\cal C}[D_2]\Big) {\cal C}[A]{\cal C}[B]\,.~~~\label{Y21}
\eea
Now we fix $A$, $B$, and count the coefficient of ${\cal
C}[\O\Lambda_1]{\cal C}[\O\Lambda_2]{\cal C}[\O\Lambda_3]$ with
$n\in \O\Lambda_1$ under the summation over divisions $\Spaa{CD}$.
The correct configurations have been given when considering $Y_1$,
thus the coefficient is
\bea & &-\Big(2p_{\O\Lambda_1\O\Lambda_2}\cdot
p_{\O\Lambda_3}+4p_{\O\Lambda_3}\cdot
p_B+2p_{\O\Lambda_1\O\Lambda_3}\cdot
p_{\O\Lambda_2}+4p_{\O\Lambda_2}\cdot p_B
+2p_{\O\Lambda_2\O\Lambda_3}\cdot
p_{\O\Lambda_1}+4p_{\O\Lambda_1}\cdot p_B\Big)\nn
&=&-2(s_A-s_B)+2s_{\O\Lambda_1}+2s_{\O\Lambda_2}+2s_{\O\Lambda_3}\,.
\eea
Substituting it into \eref{Y21} we get
\bea Y_2&=&\sum_{1,m\in A\subset
\Lambda}~\sum_{\Spaa{\O\Lambda_1\O\Lambda_2\O\Lambda_3}}~\Big(-2(s_A-s_B)+2s_{\O\Lambda_1}+2s_{\O\Lambda_2}+2s_{\O\Lambda_3}\Big)
\big({\cal C}[A]{\cal C}[B]\big)\big({\cal C}[\O\Lambda_1]{\cal
C}[\O\Lambda_2]{\cal C}[\O\Lambda_3]\big)\nn
&=&-2\sum_{\Spaa{\Lambda_1\Lambda_2\Lambda_3}}\sum_{\Spaa{\O\Lambda_1\O\Lambda_2\O\Lambda_3}}
~~\big({\cal C}[\Lambda_1]{\cal C}[\Lambda_2]{\cal
C}[\Lambda_3]\big) \big({\cal C}[\O\Lambda_1]{\cal
C}[\O\Lambda_2]{\cal C}[\O\Lambda_3]\big)\nn & &+2\sum_{1,m\in
A\subset \Lambda}~{\cal C}[A]{\cal
C}[B]~\sum_{\Spaa{\O\Lambda_1\O\Lambda_2\O\Lambda_3}}~\Big(\sum_{\Spaa{\O\Lambda_{11}\O\Lambda_{12}}}{\cal
C}[\O\Lambda_{11}]{\cal C}[\O\Lambda_{12}]{\cal C}[\O\Lambda_2]{\cal
C}[\O\Lambda_3] +\sum_{\Spaa{\O\Lambda_{21}\O\Lambda_{22}}}{\cal
C}[\O\Lambda_1]{\cal C}[\O\Lambda_{21}]{\cal C}[\O\Lambda_{22}]{\cal
C}[\O\Lambda_3]\nn &
&+\sum_{\Spaa{\O\Lambda_{31}\O\Lambda_{32}}}{\cal
C}[\O\Lambda_1]{\cal C}[\O\Lambda_2]{\cal C}[\O\Lambda_{31}]{\cal
C}[\O\Lambda_{32}]\Big)\,, \eea
where \eref{ccc} has been used again. In the last line,
$\O\Lambda_{i1}$ and $\O\Lambda_{i2}$ are two simple subsets of
$\O\Lambda_i$. For the term ${\cal C}[\lambda_1]{\cal
C}[\lambda_2]{\cal C}[\lambda_3]{\cal C}[\lambda_4]$ with $n\in
\lambda_1$ and
$\lambda_1\bigcup\lambda_2\bigcup\lambda_3\bigcup\lambda_4=\O\Lambda$, its
coefficient under the summation over divisions
$\Spaa{\O\Lambda_1\O\Lambda_2\O\Lambda_3}$ can be counted by
following six configurations:
$\{\O\Lambda_{11}=\lambda_1,\O\Lambda_{12}=\lambda_2,\O\Lambda_2=\lambda_3,\O\Lambda_3=\lambda_4\}$,
$\{\O\Lambda_{11}=\lambda_1,\O\Lambda_{12}=\lambda_3,\O\Lambda_2=\lambda_2,\O\Lambda_3=\lambda_4\}$,
$\{\O\Lambda_{11}=\lambda_1,\O\Lambda_{12}=\lambda_4,\O\Lambda_2=\lambda_2,\O\Lambda_3=\lambda_3\}$,
$\{\O\Lambda_1=\lambda_1,\O\Lambda_{21}=\lambda_2,\O\Lambda_{22}=\lambda_3,\O\Lambda_3=\lambda_4\}$,
$\{\O\Lambda_1=\lambda_1,\O\Lambda_{21}=\lambda_2,\O\Lambda_{22}=\lambda_4,\O\Lambda_3=\lambda_3\}$,
$\{\O\Lambda_1=\lambda_1,\O\Lambda_{21}=\lambda_3,\O\Lambda_{22}=\lambda_4,\O\Lambda_3=\lambda_2\}$.
Thus, we find
\bea
Y_2&=&-2\sum_{\Spaa{\Lambda_1\Lambda_2\Lambda_3}}\sum_{\Spaa{\O\Lambda_1\O\Lambda_2\O\Lambda_3}}
~~\big({\cal C}[\Lambda_1]{\cal C}[\Lambda_2]{\cal
C}[\Lambda_3]\big) \big({\cal C}[\O\Lambda_1]{\cal
C}[\O\Lambda_2]{\cal C}[\O\Lambda_3]\big)\nn & &+12\sum_{1,m\in
A\subset \Lambda}~{\cal C}[A]{\cal
C}[B]~\sum_{\Spaa{\lambda_1\lambda_2\lambda_3\lambda_4}}~{\cal
C}[\lambda_1]{\cal C}[\lambda_2]{\cal C}[\lambda_3]{\cal
C}[\lambda_4]\,. \eea
Similar to the appearing of quartic vertexes, the term ${\cal
C}[\lambda_1]{\cal C}[\lambda_2]{\cal C}[\lambda_3]{\cal
C}[\lambda_4]$ can be explained as a new quintic vertex.

Finally, we consider $Y_3$ to obtain
\bea Y_3=-4\sum_{1,m\in A\subset \Lambda}~{\cal C}[A]{\cal
C}[B]~\sum_{n\in D\subset\O \Lambda}~\Big(\sum_{\Spaa{C_1C_2}}{\cal C}[C_1]{\cal C}[C_2]\Big)\Big(\sum_{\Spaa{D_1D_2}}{\cal
C}[D_1]{\cal C}[D_2]\Big)\,, \eea
where $C_1$ and $C_2$ are two simple subsets of $C$, $D_1$ and $D_2$
are two simple subsets of $D$ with $n\in D_1$. Considering the
coefficient of ${\cal C}[\lambda_1]{\cal C}[\lambda_2]{\cal
C}[\lambda_3]{\cal C}[\lambda_4]$ with $n\in \lambda_1$ under the
summation over divisions $\Spaa{CD}$, one can find three correct
configurations:
$\{D_1=\lambda_1,D_2=\lambda_2,C_1=\lambda_3,C_2=\lambda_4\}$,
$\{D_1=\lambda_1,D_2=\lambda_3,C_1=\lambda_2,C_2=\lambda_4\}$ and
$\{D_1=\lambda_1,D_2=\lambda_4,C_1=\lambda_2,C_2=\lambda_3\}$, thus
we arrive at
\bea Y_3=-12\sum_{1,m\in A\subset \Lambda}~{\cal C}[A]{\cal
C}[B]~\sum_{\Spaa{\lambda_1\lambda_2\lambda_3\lambda_4}}~{\cal
C}[\lambda_1]{\cal C}[\lambda_2]{\cal C}[\lambda_3]{\cal
C}[\lambda_4]\,. \eea

Adding $Y_1$, $Y_2$ and $Y_3$ together, we can see that the quintic vertexes cancels each other,
and the result is given as
\bea \sum_{1,m\in A\subset \Lambda}~~\sum_{n\in D\subset\O
\Lambda}~~{\cal K}_1{\cal C}[A]{\cal C}[B]{\cal C}[C]{\cal C}[D]
=\sum_{\Spaa{\Lambda_1\Lambda_2\Lambda_3}}\sum_{\Spaa{\O\Lambda_1\O\Lambda_2\O\Lambda_3}}
~~\big({\cal C}[\Lambda_1]{\cal C}[\Lambda_2]{\cal
C}[\Lambda_3]\big) \big({\cal C}[\O\Lambda_1]{\cal
C}[\O\Lambda_2]{\cal C}[\O\Lambda_3]\big)\,. \eea
Similar calculation gives
\bea \sum_{1\in A\subset \Lambda}~~\sum_{n,m+1\in D\subset\O
\Lambda}~~{\cal K}_2{\cal C}[A]{\cal C}[B]{\cal C}[C]{\cal C}[D]
&=&\sum_{\Spaa{\Lambda_1\Lambda_2\Lambda_3}}\sum_{\Spaa{\O\Lambda_1\O\Lambda_2\O\Lambda_3}}
~~\big({\cal C}[\Lambda_1]{\cal C}[\Lambda_2]{\cal
C}[\Lambda_3]\big) \big({\cal C}[\O\Lambda_1]{\cal
C}[\O\Lambda_2]{\cal C}[\O\Lambda_3]\big)\,.~~~~~~ \eea

To finish our proof, let us we consider the last term of relation
\eref{rela-3}
\bea & &\sum_{1\in A\subset \Lambda}~~\sum_{n\in D\subset\O
\Lambda}~~ {1\over 9}(s_A+s_B)(s_C+s_D){\cal C}[A]{\cal C}[B]{\cal
C}[C]{\cal C}[D]\nn &=&\sum_{1\in A\subset \Lambda}~~\sum_{n\in
D\subset\O \Lambda}~~ {1\over 9}\Big(\sum_{\Spaa{A_1A_2}}{\cal
C}[A_1]{\cal C}[A_2]{\cal C}[B] +\sum_{\Spaa{B_1B_2}}{\cal
C}[A]{\cal C}[B_1]{\cal C}[B_2]\Big)\nn &
&\Big(\sum_{\Spaa{C_1C_2}}{\cal C}[C_1]{\cal C}[C_2]{\cal C}[D]
+\sum_{\Spaa{D_1D_2}}{\cal C}[C]{\cal C}[D_1]{\cal C}[D_2]\Big)\,,
\eea
with $1\in A_1$, $n\in D_1$. This situation is familiar for us now.
There are three configurations for ${\cal C}[\Lambda_1]{\cal
C}[\Lambda_2]{\cal C}[\Lambda_3]$ with $1\in \Lambda_1$ and three
configurations for ${\cal C}[\O\Lambda_1]{\cal C}[\O\Lambda_2]{\cal
C}[\O\Lambda_3]$ with $n\in \O\Lambda_1$, thus there are totally
nine configuations for the term ${\cal C}[\Lambda_1]{\cal
C}[\Lambda_2]{\cal C}[\Lambda_3]{\cal C}[\O\Lambda_1]{\cal
C}[\O\Lambda_2]{\cal C}[\O\Lambda_3]$ which leads
\bea & &\sum_{1\in A\subset \Lambda}~~\sum_{n\in D\subset\O
\Lambda}~~ {1\over 9}(s_A+s_B)(s_C+s_D){\cal C}[A]{\cal C}[B]{\cal
C}[C]{\cal C}[D]\nn
&=&\sum_{\Spaa{\Lambda_1\Lambda_2\Lambda_3}}\sum_{\Spaa{\O\Lambda_1\O\Lambda_2\O\Lambda_3}}
~~\big({\cal C}[\Lambda_1]{\cal C}[\Lambda_2]{\cal
C}[\Lambda_3]\big) \big({\cal C}[\O\Lambda_1]{\cal
C}[\O\Lambda_2]{\cal C}[\O\Lambda_3]\big)\,. \eea
Thus, we have
\bea & & \sum_{1,m\in A\subset \Lambda}~~\sum_{n\in D\subset\O
\Lambda}~~{\cal K}_1{\cal C}[A]{\cal C}[B]{\cal C}[C]{\cal C}[D]\nn
&=&\sum_{1\in A\subset \Lambda}~~\sum_{n,(m+1)\in D\subset\O
\Lambda}~~{\cal K}_2{\cal C}[A]{\cal C}[B]{\cal C}[C]{\cal C}[D]\nn
&=&\sum_{1\in A\subset \Lambda}~~\sum_{n\in D\subset\O \Lambda}~~
{1\over 9}(s_A+s_B)(s_C+s_D){\cal C}[A]{\cal C}[B]{\cal C}[C]{\cal
C}[D]\nn
&=&\sum_{\Spaa{\Lambda_1\Lambda_2\Lambda_3}}\sum_{\Spaa{\O\Lambda_1\O\Lambda_2\O\Lambda_3}}
~~\big({\cal C}[\Lambda_1]{\cal C}[\Lambda_2]{\cal
C}[\Lambda_3]\big) \big({\cal C}[\O\Lambda_1]{\cal
C}[\O\Lambda_2]{\cal C}[\O\Lambda_3]\big)\,. \eea
This ends the proof of relation \eref{rela-3}.

Before ending this section, we would like to point out a by-product in this
section. We have shown that the final result of the integration can
be expressed as
\bea {\cal A}&=&{1\over s_{\Lambda}^3}\Big(\sum_{\Spaa{\Lambda\to
AB}}\sum_{\Spaa{\O\Lambda\to CD}}~\O X{\cal C}[A]{\cal C}[B]{\cal
C}[C]{\cal C}[D] -{1\over 4}s_{\Lambda}\sum_{\Spaa{\O\Lambda\to
CD}}~\sum_{\Spaa{\Lambda\to
\Lambda_1\Lambda_2\Lambda_3}}~\big({\cal C}[C]{\cal
C}[D]\big)\big({\cal C}[\Lambda_1]{\cal C}[\Lambda_2]{\cal
C}[\Lambda_3]\big)\nn & &-{1\over
4}s_{\O{\Lambda}}\sum_{\Spaa{\Lambda\to
AB}}~\sum_{\Spaa{\O\Lambda\to
\O\Lambda_1\O\Lambda_2\O\Lambda_3}}~\big({\cal C}[A]{\cal
C}[B]\big)\big({\cal C}[\O\Lambda_1]{\cal C}[\O\Lambda_2]{\cal
C}[\O\Lambda_3]\big)\nn & &+{1\over
4}\sum_{\Spaa{\Lambda\to\Lambda_1\Lambda_2\Lambda_3}}\sum_{\Spaa{
\O\Lambda\to\O\Lambda_1\O\Lambda_2\O\Lambda_3}} ~~\big({\cal
C}[\Lambda_1]{\cal C}[\Lambda_2]{\cal C}[\Lambda_3]\big) \big({\cal
C}[\O\Lambda_1]{\cal C}[\O\Lambda_2]{\cal
C}[\O\Lambda_3]\big)\Big)\,, \eea
where $\O X$ is defined in \eref{xbar}. This formulation can be considered as a new kind of integration rule.
Now instead of just cubic vertexes like $\Spaa{\Lambda\to AB}$ and
$\Spaa{\O\Lambda\to CD}$, there are new quartic  vertexes like
$\Spaa{\Lambda\to \Lambda_1\Lambda_2\Lambda_3}$ and
$\Spaa{\O\Lambda\to \O\Lambda_1\O\Lambda_2\O\Lambda_3}$.

\section{Rule for duplex-double pole}
\label{secrule3}

In this section, we will derive the Feynman rule for duplex-double pole. In \cite{Huang:2016zzb},
the Feynman rule is conjectured to be
\bea
\ruleIII{p_A,p_B,p_E,p_C,p_D}&=&{(2p_A\cdot p_D)(2p_B\cdot p_C)-(2p_A\cdot p_C)(2p_B\cdot p_D)\over
s_{AB}^2s_{CD}^2}\nonumber\\
&&-{(p_E^2)(2p_A\cdot p_D+2p_B\cdot p_C-2p_A\cdot p_C-2p_B\cdot p_D)\over
4s_{AB}^2s_{CD}^2}\,.~~~\label{rule3}\eea
In this section, we will derive a different rule for the special case $p_E^2=0$ for simplicity. The new rule will be
different from the one given in \eref{rule3}. Since we have restricted ourselves only on the special case,
we will not present the equivalent proof of these two
rules in this paper.

\subsection{CHY configuration}

\begin{figure}[htb]
\centering \subfigure[Original integrand]{
\label{fig:rule3:A} 
\includegraphics[scale=0.5]{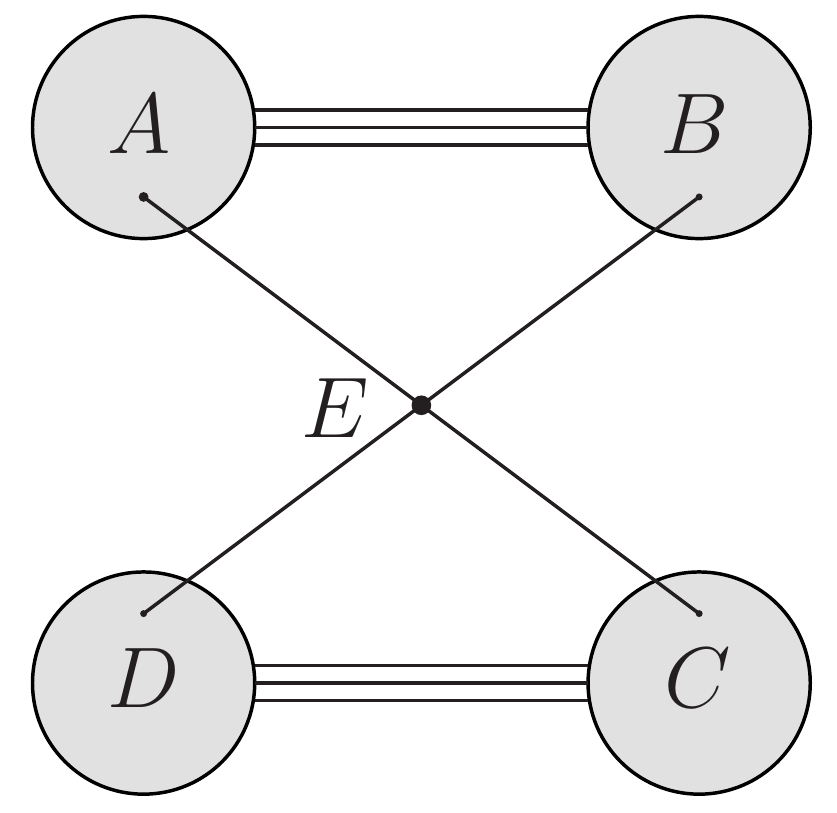}}%
\hfill%
\subfigure[Cross-ratio identity: the $T_1$ part]{
\label{fig:rule3:B} 
\includegraphics[scale=0.5]{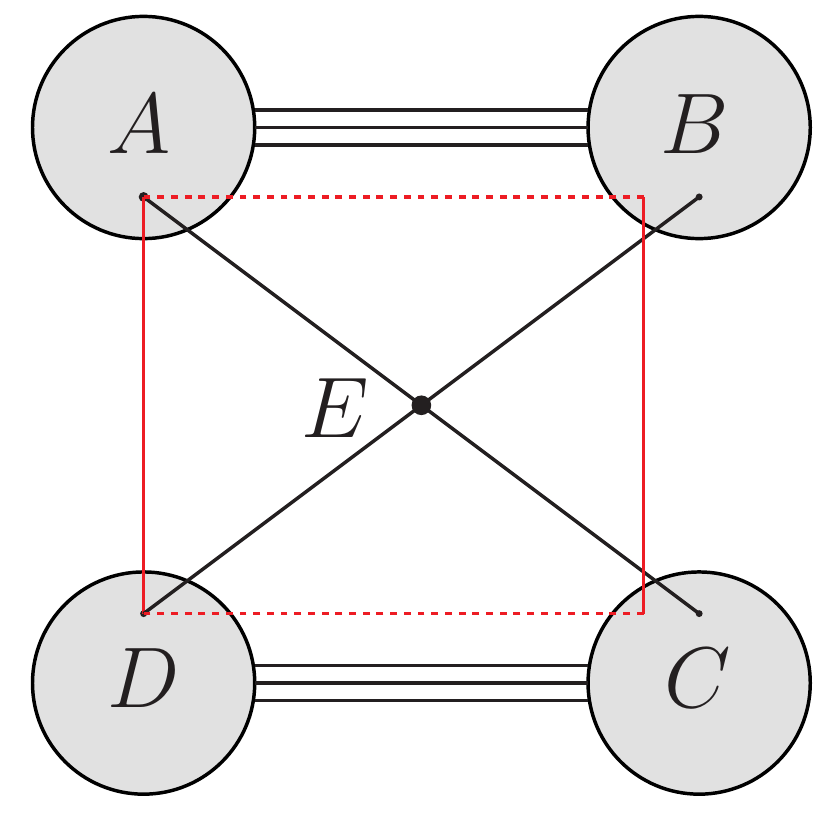}}
\hfill%
\subfigure[Cross-ratio identity: the $T_2$ part]{
\label{fig:rule3:C} 
\includegraphics[scale=0.5]{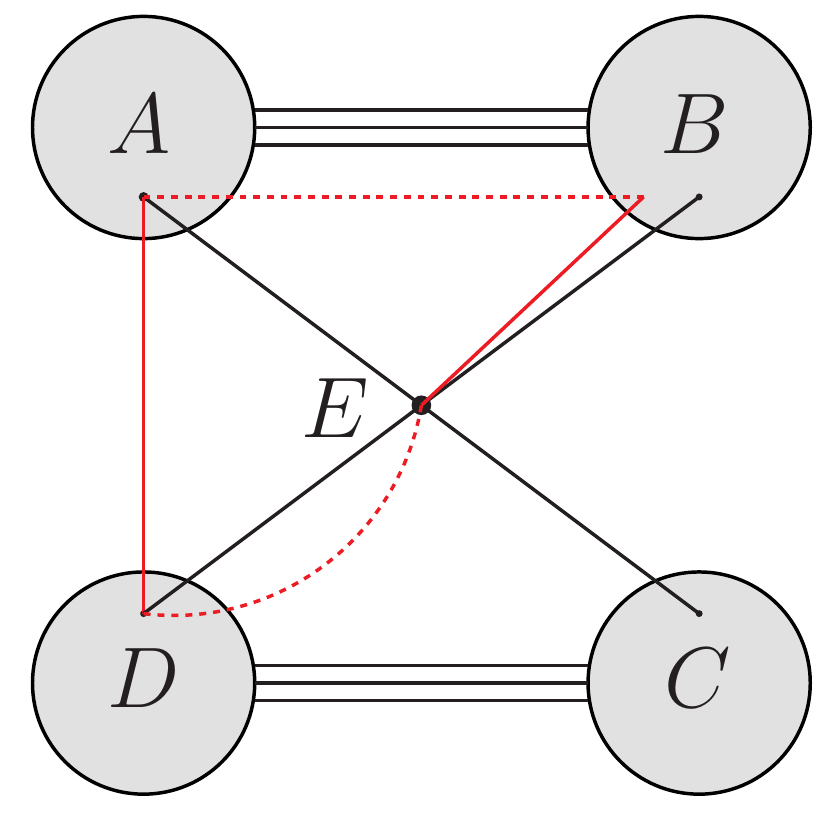}}
\caption{ Configuration of the original integrand and the cross-ratio identity of the rule III. The red lines represent terms
${z_{jd}z_{ia}\over z_{ij}z_{ad}}$ and ${z_{ed}z_{ia}\over z_{ie}z_{ad}}$
provided by the cross-ratio identity. }
\label{fig:rule3} 
\end{figure}

Like other two cases, first let us specify the CHY configurations in this section, i.e., all poles are simple poles except two double poles
$s_{\Lambda_1}$ and $s_{\Lambda_2}$ ($\Lambda_1\bigcap\Lambda_2=\emptyset$). Furthermore, to simplify the problem, we have assumed the numerator is one for CHY integrands and  the $E=\O {\Lambda_1\bigcup \Lambda_2}$ is just a single node $e$.
With above specifications, we can derive the following statements:
\begin{itemize}

\item (1) By the same argument  for the case with only one double pole, for the subset $\Lambda_1$, there are two points $a, b$ having a line connecting to subset $\O \Lambda_1$. Similarly, for the subset $\Lambda_2$, there are two points $c, d$ having a line connecting to subset $\O \Lambda_2$.

\item (2) Because    $E=\{e\}$, we see immediately that
there are four lines $[a e]$, $[b e]$,  $[c e]$ and  $[d e]$.

\item (3) For all Feynman diagrams, $\Lambda_1$ will split into two simple corners $A$ and $B$ with $a\in A$, $b\in B$,
and similarly $\Lambda_2$ will split into two simple corners $C$ and $D$ with $c\in C$,
$d\in D$, as can be seen in Figure \ref{fig:rule3:A}.

\item (4) Using \eref{exclude-Sigma},  there is no single pole
$\Sigma=\a\bigcup\b$ with  true subsets $\a \subset  \Lambda_1$, $\b\subset
 \Lambda_2$ by the same argument.

\item (5) Using \eref{exclude-Sigma},  there is no single pole
$\Sigma=\a\bigcup E$ with the true subset  $\a \subset  \Lambda_1$. The reason is that
if $\chi[\a]=0$, then $\mathbb{L}[\a,E]\leq 1$ since $b, a$ can not belong to same single pole.
In other words, we will have $\mathbb{L}[\a,E]+\chi[\a]\leq 1$.
Similarly there is no single pole
$\Sigma=\b\bigcup E$ with the true subset  $\b \subset  \Lambda_2$.

\item (6) For the case $\Sigma=\a\bigcup\b\bigcup E$ with  true subsets $\a \subset  \Lambda_1$, $\b\subset
 \Lambda_2$, we find
 \bea \chi[\Sigma]=\chi[\a]+\chi[\b]+\mathbb{L}[\a,\b]+\mathbb{L}[\a,E]+\mathbb{L}[\b,E]-4\,,~~~\label{3-subsets}\eea
with $\mathbb{L}[\a,\b]=0$, $\mathbb{L}[\a,E]+\chi[\a]\leq 1$ and $\mathbb{L}[\b,E]+\chi[\b]\leq 1$, we get  $\chi[\Sigma]<0$ always, i.e., there is no such a simple pole.

\end{itemize}

Based on above observations, we see that all maximal compatible
combinations must contain $\Lambda_1$ and $\Lambda_2$. The reason is
that there are  at most $(|\Lambda_1|-2)$ compatible combinations
from $\Lambda_1$ and $(|\Lambda_2|-2)$ compatible combinations from
$\Lambda_2$, thus at most we can get  $(n-5)$ compatible poles.
Therefore one must add two poles $s_{\Lambda_1}$ and $s_{\Lambda_2}$
to achieve the correct number $(n-3)$ for all poles. In other words,
all allowed Feynman diagrams will contain the following cubic vertex
where double poles  $s_{\Lambda_1}$ and $s_{\Lambda_2}$ meet with
the node $e$. With this picture, the integrated result should
be\footnote{For single note, since ${\cal C}[\{e\}]=1$, one can drop
this factor. }
\bea I_{CHY} & \to &  \sum_{\Spaa{ABCD}}{X\over
s_{\Lambda_1}^2s_{\Lambda_2}^2}{\cal C}[A] {\cal C}[B]{\cal
C}[C]{\cal C}[D]{\cal C}[E]\,, \nn
& & a\in A\subset\Lambda_1,~b\in B=\Lambda_1/A,~c\in \Lambda_2/D,~d\in D\subset\Lambda_2\,.~~~~
\label{rule3-pole}\eea
Now we need to determine the expression $X$ to get the Feynman rule.

\subsection{Derivation}

Now we derive the rule. Since there are two double poles $s_{\Lambda_1}$ and $s_{\Lambda_2}$,
we need to  use the cross-ratio identities twice. In the first step, our
choice of the gauge is $[a,d,{\Lambda_1}]$, then the identity is given by
\bea -1=\sum_{i\in \Lambda_1\setminus\{a\}}\sum_{j\in
{\Lambda_2}\setminus\{d\}} {s_{ij}\over
s_{\Lambda_1}}{z_{jd}z_{ia}\over z_{ij}z_{ad}}+ \sum_{i\in
\Lambda_1\setminus\{a\}}{s_{ie}\over
s_{\Lambda_1}}{z_{ed}z_{ia}\over z_{ie}z_{ad}}\,,~~~~\label{iden3}
\eea
where we have split the sum $j\in \O\Lambda_1/\{d\}$ into two parts,
as shown in Figure \ref{fig:rule3:B} and Figure \ref{fig:rule3:C} respectively.
For the first part, the factor ${z_{jd}z_{ia}\over z_{ij}z_{ad}}$
has reduced the double pole $s^2_{\Lambda_2}$ to simple pole
$s_{\Lambda_2}$ simultaneously by the numerator $z_{jd}$.
Furthermore, it does not create any new simple poles. Firstly, it can
not create new simple poles of the form $\a\bigcup E$ with $\a$ the
true subsets of either $\Lambda_1$ or $\Lambda_2$. Secondly,  it can
not create new simple poles of the form $\a\bigcup\b$ with $\a,\b$ the
true subsets of $\Lambda_1$ and $\Lambda_2$ respectively. Using
\eref{exclude-Sigma}, one see that after multiplying the factor
${z_{jd}z_{ia}\over z_{ij}z_{ad}}$, $L[\a,\b]$ can change from zero
to one at most. Similarly,  it can not create new simple poles of the
form  $\a\bigcup \b\bigcup E$ with $\a,\b$ the true subsets of
$\Lambda_1$ and $\Lambda_2$ respectively. For this one, we need to
use  \eref{3-subsets} and again $L[\a,\b]$ can change from zero to
one at most. Above analysis shows that the first term has been
reduced to the case without any higher order poles. Thus the integration
rules given in \cite{Baadsgaard:2015voa,
Baadsgaard:2015ifa,Baadsgaard:2015hia} can be applied
straightforwardly. Putting all considerations together, the first part
of \eref{iden3} gives
\bea T_1=\sum_{\Spaa{ABCD}}{2p_B\cdot p_C\over s_{\Lambda_1}}{1\over
s_{\Lambda_1}s_{\Lambda_2}}{\cal C}[A] {\cal C}[B]{\cal C}[C]{\cal
C}[D]{\cal C}[E]\,.~~~\label{Rule3-Part-1}\eea

Now we consider the second part of \eref{iden3}. Because of the factor
${z_{ed}z_{ia}\over z_{ie}z_{ad}}$, especially the denominator
$z_{ie}$, it is easy to see from \eref{exclude-Sigma} that now the
subset $\a\bigcup E$ with $i,b\in \a, a\not\in \a$, $\a\subset
\Lambda_1$ and $\chi[\a]=0$ will become a new single pole. This
phenomenon will complicate our discussion a lot. Also, for this
part, the double pole $s^2_{\Lambda_2}$ still exists, thus we need
to use cross ratio identity again. In fact, if we set
$\W\Lambda_1=\Lambda\bigcup E=\O \Lambda_2$ and $\W
\Lambda_2=\Lambda_2$, $I_{org}{z_{ed}z_{ia}\over z_{ie}z_{ad}}$ is
the case with only one double pole $s_{\Lambda_2}$ studied in
previous section, where $a,e$ are special points in $\W\Lambda_1$
and $d,c$ are the special points in $\W\Lambda_2$. Using result
\eref{rule-I-Feyn-2} we get
\bea { (2p_B\cdot p_E)\over s_{\Lambda_1}} \sum_{\W A\subset
\W\Lambda_1} \sum_{C\subset \Lambda_2} { (2 p_{\W A}\cdot p_C) \over
s^2_{\Lambda_2}}{\cal C}[\W A] {\cal C}[\W B]{\cal C}[C]{\cal
C}[D]\,,\eea
where we have used the fact $\W \Lambda_2=\Lambda_2$. Now coming to
the key point: the allowed splitting of $\W\Lambda_1$ can be divided
into two cases. In the first case, $\W B=E$ and $\W
A=\Lambda_1$. Furthermore, the $\W A$ split into two corners $A,B$
and we have
\bea {\cal C}[\W A] {\cal C}[\W B]= \left( {1\over s_{\Lambda_1}}
\sum_{\Spaa{AB}}{\cal C}[A]{\cal C}[B]\right){\cal C}[E]\,.\eea
Putting it back, we get
\bea T_{2;1} & = & \sum_{\Spaa{ABCD}}{ (2p_B\cdot p_E)\over
s_{\Lambda_1}}  { (2 (p_A+p_B)\cdot p_C) \over
s^2_{\Lambda_2}}{1\over s_{\Lambda_1}} {\cal C}[A]{\cal C}[B]{\cal
C}[C]{\cal C}[D]{\cal C}[E]\,.~~~\label{part2}\eea
In the second case, $\W A=A$ and $\W B=B \bigcup E$. Unlike the first case, we know for sure that ${\cal C}[\W
A]= \left( {1\over s_{\Lambda_1}}\sum_{\Spaa{AB}} {\cal C}[A]{\cal C}[B]\right)$.
Here node $e$ can be attached to some propagators in the
sub-Feynamn diagrams of subset $B$. We will show that the final
result for the second case is
\bea T_{2;2} & = & \sum_{\Spaa{ABCD}}{(2p_A \cdot
p_C)s_{\Lambda_1}\over s_{\Lambda_1}^2s_{\Lambda_2}^2}{\cal
C}[A]{\cal C}[B]{\cal C}[C]{\cal C}[D]{\cal
C}[E]\,.~~~~\label{part3} \eea
Putting \eref{Rule3-Part-1}, \eref{part2} and \eref{part3} together
with proper sign, we get the final result\footnote{The relative
minus sign is because for the second part we have inserted another
cross ratio identity.}
\bea T_1-T_{2;1}-T_{2;2}\,,\eea
which implies that the $X$ in \eref{rule3-pole} is given by
\bea X_{\Spaa{ABCD}} & = & (2p_B\cdot p_C) s_{\Lambda_2}- (2p_B\cdot
p_E) (2 (p_A+p_B)\cdot p_C)-(2p_A \cdot
p_C)s_{\Lambda_1}\,.~~~\label{X-exp}\eea
The Feynman rule \eref{rule3-pole} with $X$ given by \eref{X-exp}
has been checked numerically with several examples.

\begin{figure}[htb]
\centering \subfigure[The Feynmann diagram $\Gamma$]{
\label{fig:sequ:A} 
\includegraphics[scale=0.8]{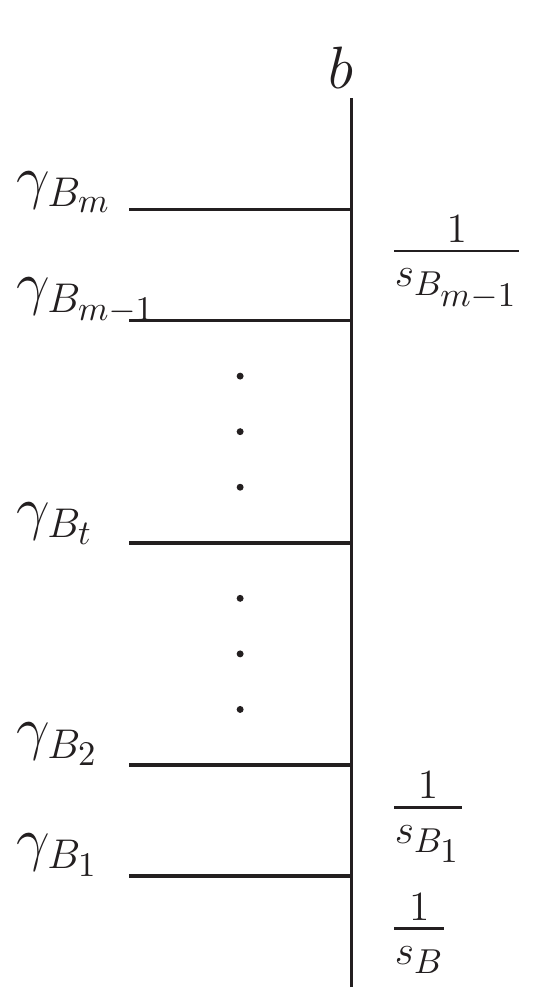}}%
\hfill%
\subfigure[$e$ attaches to the propagator ${1\over s_{B_t}}$]{
\label{fig:sequ:B} 
\includegraphics[scale=0.8]{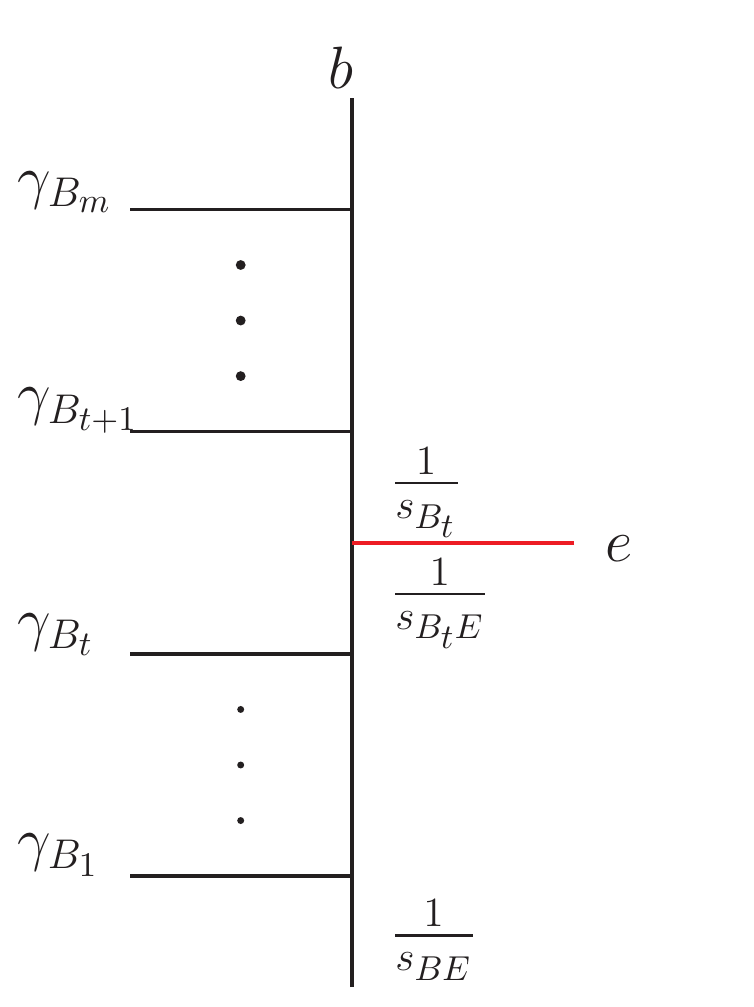}}
\caption{The Feynmann diagram $\Gamma$ of the sequence $B_m\subset B_{m-1}\subset\cdots\subset
B_1\subset B$ and the node $e$ attaches to this diagram. }
\label{fig:sequence} 
\end{figure}

Now we explain the result \eref{part3} for the second splitting of
$\W\Lambda_1=\W A\bigcup \W B$ with $\W A=A$ and $\W B=B \bigcup E$.
For a given subset $B$, there are several compatible combinations,
i.e., several sub-Feynman diagrams. Let us focus on a particular
sub-Feynman diagram $\Gamma$, which is shown in Figure \ref{fig:sequ:A}. For this $\Gamma$, there is a sequence
of single subsets such that $B_m\subset B_{m-1}\subset\cdots\subset
B_1\subset B$ with $B_m=\{b\}$. The reason we consider this sequence
is that by our previous argument, it is exactly these subsets $B_t$,
which can combine with the subset $E=\{e\}$ to create a new single
pole when the sum $i\in B_t$ in the second part of \eref{iden3}. In
other words, for the second splitting of $\W\Lambda_1$, the node $e$
will attach exactly to these propagators $B_t$ in the Feynman
diagram $\Gamma$, as can be seen in Figure \ref{fig:sequ:B}. Now we can write down the expression when node $e$
is attached to propagator $B_t$ as
\bea T_{\Gamma; B_t}  & = & \left\{ \begin{array}{ll} { 2
p_{B_t}\cdot p_E\over s_{BE} s_{B_1 E}... s_{B_t E} s_{B_t}
s_{B_{t+1}}...s_{B_{m-1}}} \gamma & ~~~t\geq m-1\,,\\ & \\
{2 p_{B_t}\cdot p_E\over s_{BE} s_{B_1 E}... s_{B_{m-1}E} s_{B_m E}}
\gamma & ~~~t=m\,.\end{array} \right.~~~~\label{Bt-exp}\eea
Let us explain the meaning of \eref{Bt-exp}. First, to be able to
attach $e$ to the propagator $B_t$, the sum $i$ in the second part
of \eref{iden3} can only be those $i\in B_t$, so we get the
numerator $2 p_{B_t}\cdot p_E$. Secondly, along the sequence of
propagators, when the node $e$ has been attached, the later
propagators will carry corresponding momentum, so we have the
propagators $s_{B_t E}$, $s_{B_{t-1} E}$, $s_{B_{t-2} E}$ until
$s_{BE}$. Thirdly, $\gamma=\prod_t \gamma_{B_t}$ is the other part
of propagators, which are not affect by node $e$, as shown in Figure
\ref{fig:sequence}. It will be the same for all $T_{\Gamma; B_t}$.

\begin{figure}[htb]
\centering
  \includegraphics[scale=0.9]{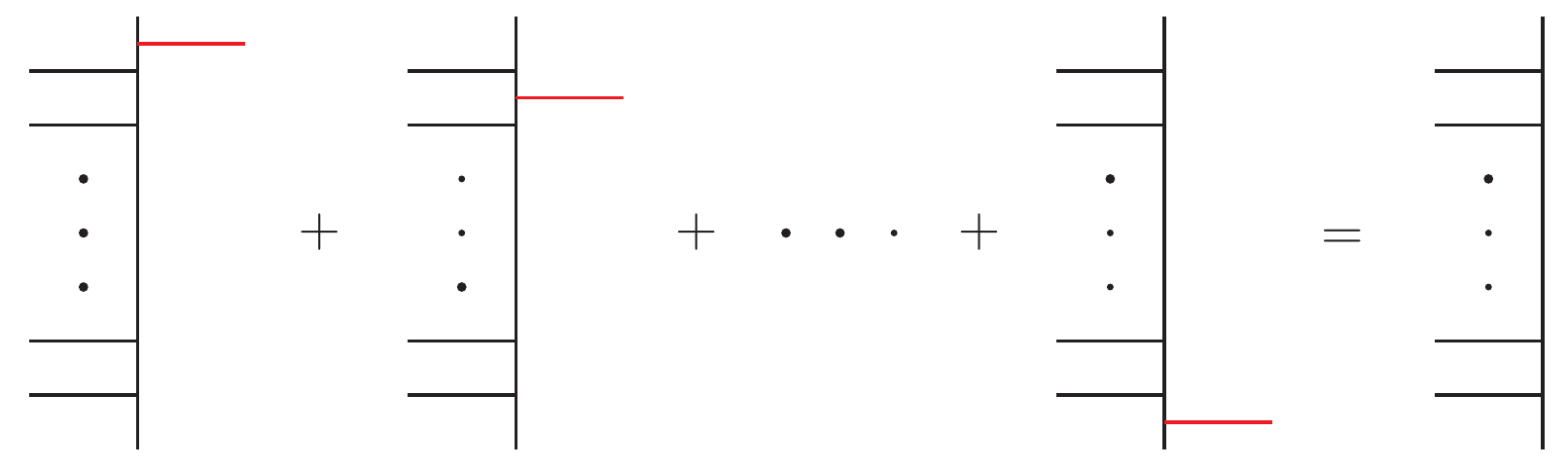}\\
  \caption{The summation $\sum_{t=0}^m  T_{\Gamma; B_{t}}$.}\label{sum}
\end{figure}

Having obtained the expression, we can carry out the sum. It is easy
to see that
\bea T_{\Gamma; B_{m-1}} + T_{\Gamma; B_{m}} & = & {\gamma\over
s_{BE} s_{B_1 E}... s_{B_{m-2}E} s_{B_{m-1}E} } \left\{
{2p_{B_{m-1}}\cdot p_E\over s_{B_{m-1}}} + {2p_{B_m}\cdot p_E\over
s_{B_m E}}\right\} \nn
& = & {\gamma\over s_{BE} s_{B_1 E}... s_{B_{m-2}E}
s_{B_{m-1}E}} \times{s_{B_{m-1}E}\over s_{B_{m-1}}}=
{\gamma\over s_{BE} s_{B_1 E}... s_{B_{m-2}E} s_{B_{m-1}}}\,, \eea
where at the second line we have used the fact $p_E^2=0$ since
$E=\{e\}$ is just a single node. Adding $T_{\Gamma; B_{m-2}}$, we
get
\bea & & T_{\Gamma; B_{m-2}}+T_{\Gamma; B_{m-1}} + T_{\Gamma;
B_{m}}\nn & = & {\gamma\over s_{BE} s_{B_1 E}... s_{B_{m-2}E}}
\left\{ {1\over s_{B_{m-1}}}+{2p_{B_{m-2}}\cdot p_E\over
s_{B_{m-2}}s_{B_{m-1}}}\right\}\nn
& = &  {\gamma\over s_{BE} s_{B_1 E}... s_{B_{m-2}E}}
\times{s_{B_{m-2}E}\over s_{B_{m-2}}s_{B_{m-1}}}= {\gamma\over
s_{BE} s_{B_1 E}... s_{B_{m-3}E} s_{B_{m-2}}s_{B_{m-1}}}\,.\eea
Now we can see the recursive pattern, which leads to
\bea \sum_{t=0}^m  T_{\Gamma; B_{t}}= {\gamma\over s_{B} s_{B_1
}... s_{B_{m-3}} s_{B_{m-2}}s_{B_{m-1}}}\,.~~~\label{T-CB-part}\eea
This is just the expression of $\Gamma$ itself. This summation can
be understood diagrammatically as in Figure \ref{sum}. This
calculation has also shown how the new created poles have been
canceled out when summing over all terms.

Having above preparations, we are ready to write down the
contribution from the second splitting of $\W\Lambda_1$ as
\bea T_{2;2}={1\over s_{\Lambda_1}}\sum_{\W\Lambda_1=A\bigcup \W B}
{2 p_A\cdot p_C\over s^2_{\Lambda_2}} {\cal C}[A] {\cal C}[C] {\cal
C}[D] \sum_{\Gamma}\left\{ \sum_{t=0}^m  T_{\Gamma; B_{t}}\right\}\,,
\eea
where the Feynman rule \eref{rule-I-Feyn-2} has been used. Using
\eref{T-CB-part} to above expression, we arrive at the wanted result
\eref{part3}.

\section{Conclusion}
\label{secconclu}

In this paper, we use the cross-ratio identity to derive the Feynman rules of higher-order poles in the CHY construction conjectured in \cite{Huang:2016zzb}. The first rule is valid for CHY integrands containing a double pole, the second rule is valid
for integrands containing a triple pole, while the third rule
is valid for integrands containing a duplex-double pole. The new expression of rules obtained in this paper depends on the gauge choices of cross-ratio
identities, however the final results of integrations after summing over all correct compatible
combinations are gauge invariant.

For the first and second rule, we make the comparison of them with those conjectured in
\cite{Huang:2016zzb}. For the second rule, the equivalence between our formula
and the conjectured one is non-trivial. We have found that the result
of integration can be arranged into a new version of rule which contains quartic
vertexes. This is an interesting phenomenon.

For the first two rules, we have performed special gauge choices to ensure that
new poles which do not satisfy the desired compatible combination
will not emerge. For the third rule however, this idea can not be realized, i.e.,
one can not avoid the appearing of new poles. Thus we have provided a treatment
of new poles for this rule.

In general, searching for integration rules for higher-order poles is not
efficient enough for practical computation, since one can not reduce the most
general CHY integrands which contain higher order poles into a few configurations.
Rules in this paper or rules
in \cite{Huang:2016zzb} only cover some special cases. When encountering new cases,
we need new rules. However, since we have proven the conjectured rules through the
cross-ratio identities, we can conclude that the cross-ratio identity method
is a powerful and universal tool for analytic calculation.

\section*{Acknowledgments}

We would thank Rijun Huang for useful discussions. This work is
supported by Qiu-Shi Funding and Chinese NSF funding under contracts
No.11575156, No.11135006, and No.11125523 as well as the USA
National Science Foundation under Grant No. NSF PHY-1125915. K.Z also
would like to acknowledge the support from Chinese Postdoctoral
Administrative Committee. B.F would like to thank the workshop
"Scattering Amplitudes and Beyond" at Kavli Institute for the
Theoretical Physics, where the final draft has been edited.



\begin{thebibliography}{}

\bibitem{Cachazo:2013gna}
  F.~Cachazo, S.~He and E.~Y.~Yuan,
  Phys.\ Rev.\ D {\bf 90}, no. 6, 065001 (2014)
  doi:10.1103/PhysRevD.90.065001
  [arXiv:1306.6575 [hep-th]].

\bibitem{Cachazo:2013hca}
  F.~Cachazo, S.~He and E.~Y.~Yuan,
  Phys.\ Rev.\ Lett.\  {\bf 113}, no. 17, 171601 (2014)
  doi:10.1103/PhysRevLett.113.171601
  [arXiv:1307.2199 [hep-th]].

\bibitem{Cachazo:2013iea}
  F.~Cachazo, S.~He and E.~Y.~Yuan,
  JHEP {\bf 1407}, 033 (2014)
  doi:10.1007/JHEP07(2014)033
  [arXiv:1309.0885 [hep-th]].

\bibitem{Cachazo:2014nsa}
  F.~Cachazo, S.~He and E.~Y.~Yuan,
  JHEP {\bf 1501}, 121 (2015)
  doi:10.1007/JHEP01(2015)121
  [arXiv:1409.8256 [hep-th]].

\bibitem{Cachazo:2014xea}
  F.~Cachazo, S.~He and E.~Y.~Yuan,
  JHEP {\bf 1507}, 149 (2015)
  doi:10.1007/JHEP07(2015)149
  [arXiv:1412.3479 [hep-th]].

\bibitem{Kalousios:2015fya}
  C.~Kalousios,
  JHEP {\bf 1505}, 054 (2015)
  doi:10.1007/JHEP05(2015)054
  [arXiv:1502.07711 [hep-th]].

\bibitem{Cardona:2015eba}
  C.~Cardona and C.~Kalousios,
  JHEP {\bf 1601}, 178 (2016)
  doi:10.1007/JHEP01(2016)178
  [arXiv:1509.08908 [hep-th]].

\bibitem{Cardona:2015ouc}
  C.~Cardona and C.~Kalousios,
  Phys.\ Lett.\ B {\bf 756}, 180 (2016)
  doi:10.1016/j.physletb.2016.03.003
  [arXiv:1511.05915 [hep-th]].

\bibitem{Dolan:2015iln}
  L.~Dolan and P.~Goddard,
  JHEP {\bf 1610}, 149 (2016)
  doi:10.1007/JHEP10(2016)149
  [arXiv:1511.09441 [hep-th]].

\bibitem{Huang:2015yka}
  R.~Huang, J.~Rao, B.~Feng and Y.~H.~He,
  JHEP {\bf 1512}, 056 (2015)
  doi:10.1007/JHEP12(2015)056
  [arXiv:1509.04483 [hep-th]].

\bibitem{Sogaard:2015dba}
  M.~S?gaard and Y.~Zhang,
  Phys.\ Rev.\ D {\bf 93}, no. 10, 105009 (2016)
  doi:10.1103/PhysRevD.93.105009
  [arXiv:1509.08897 [hep-th]].

\bibitem{Bosma:2016ttj}
  J.~Bosma, M.~S?gaard and Y.~Zhang,
  Phys.\ Rev.\ D {\bf 94}, no. 4, 041701 (2016)
  doi:10.1103/PhysRevD.94.041701
  [arXiv:1605.08431 [hep-th]].

\bibitem{Zlotnikov:2016wtk}
  M.~Zlotnikov,
  JHEP {\bf 1608}, 143 (2016)
  doi:10.1007/JHEP08(2016)143
  [arXiv:1605.08758 [hep-th]].

\bibitem{Cachazo:2015nwa}
  F.~Cachazo and H.~Gomez,
  JHEP {\bf 1604}, 108 (2016)
  doi:10.1007/JHEP04(2016)108
  [arXiv:1505.03571 [hep-th]].

\bibitem{Gomez:2016bmv}
  H.~Gomez,
  JHEP {\bf 1606}, 101 (2016)
  doi:10.1007/JHEP06(2016)101
  [arXiv:1604.05373 [hep-th]].

\bibitem{Cardona:2016bpi}
  C.~Cardona and H.~Gomez,
  JHEP {\bf 1606} (2016) 094
  doi:10.1007/JHEP06(2016)094
  [arXiv:1605.01446 [hep-th]].

\bibitem{Baadsgaard:2015voa}
  C.~Baadsgaard, N.~E.~J.~Bjerrum-Bohr, J.~L.~Bourjaily and P.~H.~Damgaard,
  JHEP {\bf 1509}, 129 (2015)
  doi:10.1007/JHEP09(2015)129
  [arXiv:1506.06137 [hep-th]].

\bibitem{Baadsgaard:2015ifa}
  C.~Baadsgaard, N.~E.~J.~Bjerrum-Bohr, J.~L.~Bourjaily and P.~H.~Damgaard,
  JHEP {\bf 1509}, 136 (2015)
  doi:10.1007/JHEP09(2015)136
  [arXiv:1507.00997 [hep-th]].

\bibitem{Baadsgaard:2015hia}
  C.~Baadsgaard, N.~E.~J.~Bjerrum-Bohr, J.~L.~Bourjaily, P.~H.~Damgaard and B.~Feng,
  JHEP {\bf 1511}, 080 (2015)
  doi:10.1007/JHEP11(2015)080
  [arXiv:1508.03627 [hep-th]].

\bibitem{Huang:2016zzb}
  R.~Huang, B.~Feng, M.~x.~Luo and C.~J.~Zhu,
  JHEP {\bf 1606}, 013 (2016)
  doi:10.1007/JHEP06(2016)013
  [arXiv:1604.07314 [hep-th]].

\bibitem{Bjerrum-Bohr:2016juj}
  N.~E.~J.~Bjerrum-Bohr, J.~L.~Bourjaily, P.~H.~Damgaard and B.~Feng,
  Nucl.\ Phys.\ B {\bf 913}, 964 (2016)
  doi:10.1016/j.nuclphysb.2016.10.012
  [arXiv:1605.06501 [hep-th]].


\bibitem{Cardona:2016gon}
  C.~Cardona, B.~Feng, H.~Gomez and R.~Huang,
  JHEP {\bf 1609}, 133 (2016)
  doi:10.1007/JHEP09(2016)133
  [arXiv:1606.00670 [hep-th]].

\bibitem{Bjerrum-Bohr:2016axv}
  N.~E.~J.~Bjerrum-Bohr, J.~L.~Bourjaily, P.~H.~Damgaard and B.~Feng,
  JHEP {\bf 1609}, 094 (2016)
  doi:10.1007/JHEP09(2016)094
  [arXiv:1608.00006 [hep-th]].

\bibitem{Nandan:2016pya}
  D.~Nandan, J.~Plefka, O.~Schlotterer and C.~Wen,
  JHEP {\bf 1610}, 070 (2016)
  doi:10.1007/JHEP10(2016)070
  [arXiv:1607.05701 [hep-th]].


\bibitem{Huang:2017ydz}
  R.~Huang, Y.~J.~Du and B.~Feng,
  arXiv:1702.05840 [hep-th].

\bibitem{Zhou:2017mbe}
  K.~Zhou,
  arXiv:1703.04403 [hep-th].

\end{thebibliography}
\end{document}